%% file: paper.tex
\documentclass[a4paper,twocolumn,epjc3]{svjour3}  
\smartqed  
\RequirePackage{graphicx}
\RequirePackage{graphicx}
\RequirePackage[hang]{subfigure}
\RequirePackage{rotating}
\RequirePackage{euscript}
\RequirePackage{appendix}
\RequirePackage{float}
\RequirePackage{amsmath}
\RequirePackage{amssymb}
\usepackage[switch]{lineno}
\usepackage[noadjust]{cite}
\usepackage{atlasphysics}

\graphicspath{{figures/}}

\journalname{Eur. Phys. J. C}
\usepackage{preprintcover}
\PreprintCoverPaperTitle{Performance of the ATLAS muon trigger in \boldmath{pp} collisions at $\sqrt{s}=8$\,TeV}
\PreprintIdNumber{CERN-PH-EP-2014-154}

\PreprintJournalName{Eur. Phys. J. C}

\PreprintCoverAbstract{The performance of the ATLAS muon trigger system is evaluated with
  proton--proton collision data collected in 2012 at the Large Hadron Collider at a centre-of-mass energy
  of 8\,TeV. It is primarily evaluated using events containing a pair of muons from the decay of {\Zboson} bosons.
  The efficiency of the single-muon trigger is measured for muons with transverse momentum
  $25 < \pt < 100$\,GeV, with a statistical uncertainty of less than 0.01\,\%
  and a systematic uncertainty of 0.6\,\%.
 The $\pt$ range for efficiency determination is extended by 
  using muons from decays of {\jpsi} mesons, {\Wboson} bosons, and top quarks. The muon trigger shows highly uniform and stable performance. 
 The performance is compared to the prediction of a detailed simulation.}

\begin{document}


\title{Performance of the ATLAS muon trigger in \boldmath{pp} collisions at $\sqrt{s}=8$\,TeV}
\author{The ATLAS Collaboration}
\maketitle

\begin{abstract}

The performance of the ATLAS muon trigger system is evaluated with
  proton--proton collision data collected in 2012 at the Large Hadron Collider at a centre-of-mass energy
  of 8\,TeV. It is primarily evaluated using events containing a pair of muons from the decay of {\Zboson} bosons.
  The efficiency of the single-muon trigger is measured for muons with transverse momentum
  $25 < \pt < 100$\,GeV, with a statistical uncertainty of less than 0.01\,\%
  and a systematic uncertainty of 0.6\,\%.
 The $\pt$ range for efficiency determination is extended by 
  using muons from decays of {\jpsi} mesons, {\Wboson} bosons, and top quarks. The muon trigger shows highly uniform and stable performance. 
 The performance is compared to the prediction of a detailed simulation.
\end{abstract}

\input{introduction.tex}

\input{trigger.tex}

\input{dataset.tex}

\input{purity.tex}

\input{resolution.tex}

\input{ztap.tex}

\input{jpsitap.tex}

\input{ttandw.tex}

\input{summary.tex}

\begin{acknowledgements}
\input{Acknowledgements.tex}

\end{acknowledgements}

\bibliographystyle{atlasBibStyleWithTitle}
\bibliography{paperbib}


\onecolumn
\clearpage
\input{atlas_authlist}

\end{document}

%% file: introduction.tex
\section{Introduction}

The presence of prompt muons in the final state is a distinctive signature for many physics processes studied
in collisions of high energy protons at the LHC. These studies, which led to
 the discovery of the Higgs boson~\cite{ref:higgs_discovery_atlas,ref:higgs_discovery_cms}, include measurements of its properties,
searches for new phenomena,
as well as measurements of Standard Model processes, such as the
production of electroweak bosons and top quarks.
Therefore, a high-perfor\-mance muon trigger is essential.
In parallel, a good simulation of trigger performance is necessary.

There are many challenges in designing and implementing triggers which select $pp$ interactions with muons in the final state with high efficiency and low transverse momentum, $\pt$, thresholds in the presence of high background conditions.
The ATLAS design deploys a three-level, multi-pronged strategy with,
\begin{enumerate}
\item custom trigger electronics at Level-1, 
\item dedicated fast algorithms to reconstruct muons and estimate their parameters at Level-2, 
\item novel techniques to retain high efficiency at the event-filter while utilising offline tracking algorithms.
\end{enumerate}
The Level-2 and event-filter together are called the High Level Trigger.
In order to address a wide variety of physics topics, ATLAS has developed a suite of triggers designed to select muons.
The single-muon trigger with {\pt} threshold of 24\,GeV
is used in many physics analyses.
In addition, muon triggers in combination with electrons, jets and
missing transverse momentum, as well as moderate-{\pt} multi-muon triggers, increase
sensitivity for various physics topics which benefit from a lower {\pt} threshold.
For the $B$-physics program, various low-{\pt} multi-muon triggers are used with a special configuration that allows
a high efficiency also for non-prompt muons \footnote{Non-prompt muons
are muons which originate from the decay of a secondary particle
rather than coming directly from the primary $pp$ interaction.}.

The ATLAS experiment collected $pp$ collision data in 2012 at a centre-of-mass energy of
8\,TeV with a maximum instantaneous luminosity of $7.7 \times 10^{33}$
$\rm cm^{-2}$ $\rm s^{-1}$.
The number of interactions occurring in the same bunch crossing (called pile-up interactions)
was about 25 on average.
In this paper, the performance of the ATLAS muon trigger is evaluated, primarily using
samples containing muon pairs from {\Zboson}-boson decays.
The performance of the low-{\pt} muon trigger is evaluated
with samples containing a pair of muons from the decay of {\jpsi} mesons.
The performance for high-{\pt} muons is evaluated using events containing 
top-quarks\footnote{Unless otherwise stated CP conjugate states are
always implied.} or {\Wboson} bosons, where a {\Wboson} boson decays into a muon and neutrino.


%% file: trigger.tex
\section{Muon trigger}
\label{sect:trigger}

\subsection{ATLAS detector}

The ATLAS detector is a multi-purpose particle physics apparatus with a forward-backward symmetric cylindrical geometry
and near $4\pi$ coverage in solid angle.\footnote{ATLAS uses a
  right-handed coordinate system with its origin at the nominal
  interaction point (IP) in the centre of the detector and the
  $z$-axis along the beam pipe. The $x$-axis points from the IP to the
  centre of the LHC ring, and the $y$-axis points upward. Cylindrical
  coordinates $(r,\phi)$ are used in the transverse plane, $\phi$
  being the azimuthal angle around the beam pipe. The pseudorapidity
  is defined in terms of the polar angle $\theta$ as
  $\eta=-\ln\tan(\theta/2)$.} The detector consists of four major
sub-systems: the inner detector, electromagnetic calorimeter, 
hadronic calorimeter  and muon spectrometer. 
A detailed description of the ATLAS detector can be found in Ref.~\cite{ref:detector_paper}.
The inner detector measures tracks up to $\abseta = 2.5$
in an axial magnetic field of 2\,T using three types of sub-detectors:
a silicon pixel detector closest to the interaction point, a semiconductor tracker surrounding the pixel detector, 
and a transition radiation straw tube tracker covering $\abseta < 2.0$
as the outermost part of the inner detector. 
The calorimeter system covers the pseudorapidity range $\abseta < 4.9$
and encloses the inner detector. 
The high-granularity liquid-argon electromagnetic sampling calorimeter is divided into one barrel
($\abseta < 1.475$) and two endcap components ($1.375 < \abseta< 3.2$). 
The hadronic calorimeter is placed directly outside the
electromagnetic calorimeter. A steel/scintillator-tile calorimeter provides hadronic coverage in the range $\abseta < 1.7$.
The endcap and forward regions, spanning $1.5< \abseta < 4.9$, are instrumented with liquid-argon calorimeters. 
The calorimeters are then surrounded by the muon spectrometer.

\subsection{Muon spectrometer}

\begin{figure}
  \begin{center}
    \includegraphics[width=0.40\textwidth, angle=90]{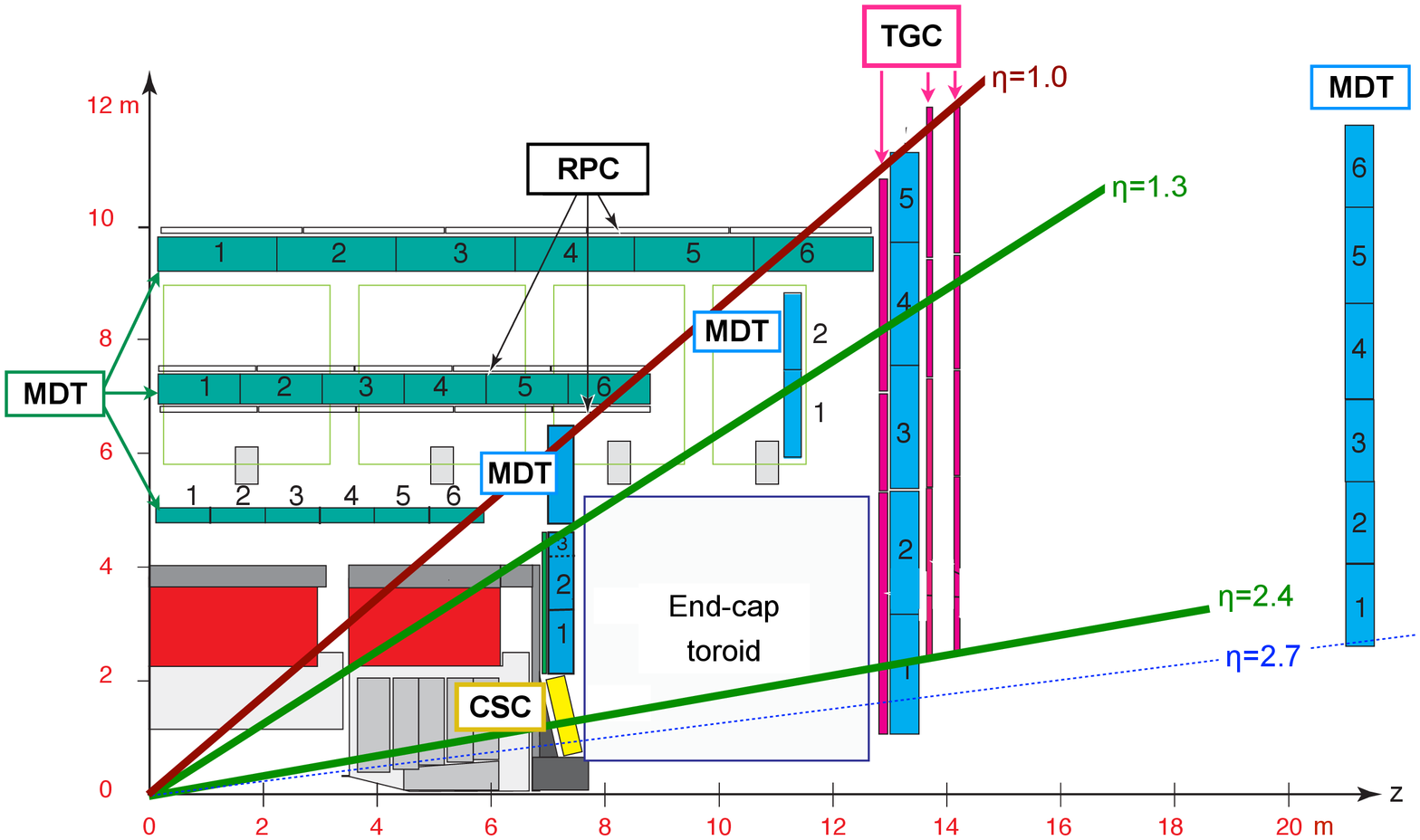}
    \caption{A schematic picture showing a quarter-section of the muon system in a plane containing the beam axis, with monitored drift tube (MDT) and cathode strip (CSC) chambers for momentum determination and resistive plate (RPC) and thin gap (TGC) chambers for triggering   
    }
    \label{fig:muon_schematic}
  \end{center}
\end{figure}

The muon spectrometer is based on three large air-core superconducting
toroidal magnet systems (two endcaps and one barrel) providing an average magnetic field of approximately 0.5\,T. 
Figure~\ref{fig:muon_schematic} shows a quarter-section of the muon
system in a plane containing the beam axis. 

In the central region, the detectors comprise a barrel that is arranged in
three concentric cylindrical shells around the beam axis.
In the endcap region, muon chambers form large wheels, perpendicular
to the $z$-axis. Several detector technologies are utilised to provide
both precision tracking and triggering.

The deflection of the muon trajectory in the magnetic field is detected using hits in three 
layers of precision monitored drift tube (MDT) chambers for $\abseta < 2$. In the region
$2.0 < \abseta < 2.7$, two layers of MDT chambers in combination with one layer of cathode strip chambers 
(CSCs) are used. Muons are independently measured in the inner detector and in the muon spectrometer.
Three layers  of resistive plate chambers (RPCs) in the barrel region ($\abseta < 1.05$),
and three layers of thin gap chambers (TGCs) in the endcap regions ($1.05 < \abseta < 2.4$) 
provide the Level-1 muon trigger.

\subsection{Level-1 muon trigger}
\label{sect:trigger_l1}

Muons are identified at Level-1 by the spatial and temporal
coincidence of hits either in the RPCs or TGCs
pointing to the beam interaction region~\cite{ref:trigger_paper,ref:detector_paper}.
The Level-1 triggers generated by hits in the RPC require a coincidence of hits in the three layers for the
highest three {\pt}  thresholds, and a coincidence of hits in two of the three layers for the rest of
thresholds. The Level-1 triggers generated by hits in the TGC require a coincidence of hits in the three layers,
except for limited areas in the lowest threshold.

The degree of deviation from the hit pattern expected for a 
muon with infinite momentum is used to estimate the {\pt} of the muon
with six possible thresholds.
The number of muon candidates passing each threshold is
used in the conditions for the global Level-1 trigger.
Following a global trigger, the {\pt} thresholds
and the corresponding detector regions, region of interest (RoIs),
are then sent to the
Level-2 and event-filter for further consideration~\cite{ref:trigger_paper,ref:detector_paper}.
The typical dimensions of the RoIs are $0.1 \times 0.1$ ($0.03 \times 0.03$) in $\Delta\eta \times \Delta\phi$ in 
the RPCs (TGCs)~\cite{ref:detector_paper}.
The geometric coverage of the 
Level-1 trigger is about 99\,\% in the endcap regions and about 80\,\% in the barrel region.
The limited geometric coverage in the barrel region is due to gaps at around $\eta = 0$ (to provide space 
for services of the inner detector and calorimeters), the feet and rib support structures of the ATLAS detector 
and two small elevator shafts in the bottom part of the spectrometer. 

\subsection{Level-2 muon trigger}

The RoI provided by Level-1
enables Level-2 to select the region of the muon detector in which the interesting features reside,
therefore reducing the amount of data to be transferred and
processed~\cite{ref:trigger_paper}.
At Level-2, a track is constructed
by adding the data from the MDT
chambers to get a more precise
estimate of the track parameters, leading to
the Level-2 stand-alone-muon~\cite{ref:cscnote_muontrigger}.
To achieve the needed resolution in sufficiently short time,
the {\pt} of the Level-2 stand-alone-muon is reconstructed with simple parameterised functions.
Then, the Level-2 stand-alone-muon is combined with a track found in the inner detector~\cite{ref:cscnote_muontrigger}.
The closest inner detector track in the $\eta$ and $\phi$ planes is selected as the best matching
track. The {\pt} value is refined by taking the weighted average between
that of the Level-2 stand-alone-muon and of the inner detector track,
leading to the so called Level-2 combined-muon.

\subsection{Event-filter muon trigger}

Muons in the event-filter are found by two different procedures.
The first focuses on RoIs defined by the Level-1 and Level-2 steps
described above and is referred to as the RoI-based method.
The second procedure searches the full detector without using the
information from the previous levels and is referred to as the full-scan method.

In the RoI-based method,
muon candidates are first formed by using the muon detectors (called event-filter stand-alone-muons), 
and are subsequently combined with inner detector tracks leading to event-filter combined-muons.
If no combined-muon is formed,
muon candidates are searched for by extrapolating inner detector tracks to the muon
detectors. If there are corresponding track segments, combined-muons are formed.
Additionally, the degree of isolation for the combined-muon is quantified
by summing the {\pt} of inner detector tracks with $\pt^{\mathrm{trk}} > 1$\,GeV found in a cone 
of $\Delta R = \sqrt{(\Delta\phi)^2 + (\Delta\eta)^2} < \Delta R_{\mathrm{cut}}$,
centred around the muon candidate after subtracting the {\pt} of the muon itself
($\Sigma_{\Delta R < \Delta R_{\mathrm{cut}}} \pt^{\mathrm{trk}}$).

The full-scan procedure is used in the event-filter to find additional muons
that are not found by the RoI-based method.
In the full-scan, muon candidates are 
first sought in the whole of the muon detectors, and then
inner detector tracks are reconstructed in the whole of the inner detectors.
Combined pairs of these inner detector and muon detector tracks form muon candidates called
event-filter full-scan-muons.

\subsection{Trigger selection criteria}
\label{sect:trigger_logic}

The trigger system is configured to use a large set of selection
criteria for each event.
Each criterion consists of
sequential selections at Level-1, Level-2  and the event-filter,
and is referred to as trigger in this paper for simplicity.
An event has to satisfy at least one of the triggers in order to be
recorded.

Table~\ref{tab:muon_chains} shows the Level-1 thresholds 
and the muon triggers discussed in this paper. 
For all trigger levels, the naming scheme typically follows a convention whereby the number that follows "mu" denotes the transverse momentum threshold and the letters, or combination of letters, characterize the muon type (isolated (i), stand alone (SA), found by full scan (FS)) and/or its origin.

The Level-1 thresholds were optimised to give an efficiency at the
designated threshold that is typically 95\,\% of the maximum
efficiency achieved well above the threshold.

%
\begin{table*}
\caption{Level-1 $\pt$ thresholds and muon triggers.
  The sequence shows the requirements at Level-1, in the event-filter or at higher level trigger which then includes Level-2.
  The requirements at Level-2 are omitted for the single- and multi-muon triggers, as they are  looser
  than those in the event-filter. The applied {\pt} and isolation requirements are also shown.
}
\label{tab:muon_chains}
\begin{center}
\begin{tabular*}{\textwidth}{@{\extracolsep{\fill}}lll@{}}
\hline
{\small Level-1} & {\small $\pt$ threshold} &{\small Number of Layers in Coincidence}\\
\hline
{\small MU4} & {\small 4\,GeV} & {\small 2 (3 in limited areas in the endcap region)} \\
{\small MU6} & {\small 6\,GeV} & {\small 2 (3 in the endcap region)} \\
{\small MU10} & {\small 10\,GeV} & {\small 2 (3 in the endcap region)} \\
{\small MU11} & {\small 10\,GeV} & {\small 3} \\
{\small MU15} & {\small 15\,GeV} & {\small 3} \\
{\small MU20} & {\small 20\,GeV} & {\small 3} \\
\hline
\end{tabular*}

\vspace{1mm}

\begin{tabular*}{\textwidth}{@{\extracolsep{\fill}}ll l@{}}
\hline
{\small Single muon trigger} & {\small Level-1} & {\small Event-Filter}\\
\hline
{\small mu6}   & {\small MU6}  & {\small One or more combined-muon with $\pt >  6$\,GeV} \\
{\small mu13}  & {\small MU10} & {\small One or more combined-muon with $\pt > 13$\,GeV} \\
{\small mu18}  & {\small MU15} & {\small One or more combined-muon with $\pt > 18$\,GeV} \\
{\small mu24i} & {\small MU15} & {\small One or more combined-muon with $\pt > 24$\,GeV and $\Sigma_{\Delta R < 0.2} \pt^{\mathrm{trk}} / \pt < 0.12$} \\
{\small mu36}  & {\small MU15} & {\small One or more combined-muon with $\pt > 36$\,GeV} \\
{\small mu40\_SA\_barrel} & {\small MU15} & {\small One or more stand-alone-muon with $\pt > 40$\,GeV in $|\eta|<1.05$}\\
\hline
\end{tabular*}

\vspace{1mm}

\begin{tabular*}{\textwidth}{@{\extracolsep{\fill}}llp{10cm}l@{}}
\hline
{\small Multi muon trigger} & {\small Level-1} & {\small Event-Filter}\\
\hline
{\small 2mu13} & {\small Two MU10} & {\small Two or more combined-muons with $\pt > 13$\,GeV (two or more mu13 triggers)} \\
{\small mu18\_mu8\_FS}  & {\small MU15} & {\small One or more combined-muon with $\pt > 18$\,GeV (mu18 trigger), and } \\
                        &               & {\small two or more full-scan muons with $\pt > 18$ and $> 8$\,GeV } \\
{\small 3mu6}  & {\small Three MU6} & {\small Three or more muons with $\pt > 6$\,GeV (three or more mu6 triggers)} \\
\hline
\end{tabular*}

\vspace{1mm}

\begin{tabular*}{\textwidth}{@{\extracolsep{\fill}}lllp{8cm}l@{}}
\hline
{\small $\jpsi$ tag-and-probe trigger} & {\small Level-1} & \multicolumn{2}{l}{\small High Level Trigger} \\
\hline
{\small mu18\_\jpsi\_FS} & {\small MU15} & {\small (Level-2)} & {\small One or more combined-muon with $\pt > 18$\,GeV} \\
                         &               & {\small (Event-filter)} & {\small One or more combined-muon with $\pt > 18$\,GeV}, and \\
                         &               &                         & {\small at least one pair of combined-muons with a mass consistent with that of $\jpsi$ }\\
{\small mu18\_\jpsi\_L2} & {\small MU15, MU4} & {\small (Level-2)} & {\small Two or more combined-muons with $\pt > 18$, $4$\,GeV}, and \\
                         &               &                         & {\small at least one pair of combined-muons with a mass consistent with that of $\jpsi$ }\\
                         &                    & {\small (Event-filter)} & {\small One or more combined-muon with $\pt > 18$\,GeV } \\
\hline
\end{tabular*}
\end{center}
\end{table*}

The triggers described in Table~\ref{tab:muon_chains} were designed
to be as inclusive as possible. 

The mu24i trigger is designed to collect isolated muons with $\pt >$ 25\,GeV with a loose isolation criterion of
$\Sigma_{\Delta R < 0.2} \pt^{\mathrm{\mathrm{trk}}} / \pt <
0.12$. The isolation criterion was chosen to retain nearly 100\%
efficiency for well isolated muons from the decays of {\Zboson}-bosons while
rejecting slightly over half of the muons from  heavy flavor,
pion and kaon decays. 

The mu36 trigger is designed to
collect muons with large {\pt} without making an isolation requirement.

The mu40\_SA\_barrel trigger is designed to recover possible inefficiency
due to muon spectrometer and inner detector combination at large {\pt}, and the
decision is based only on muon spectrometer reconstruction. It was active
only in the barrel region due to its high rate in the endcaps.

The mu24i, mu36 and mu40\_SA\_barrel triggers were used without prescale\footnote{The term prescale means that
only one in N events passing the trigger is accepted at that trigger
level, where N is an integer definite number called the prescale factor. At
Level-1 every Nth event is accepted. At the high level trigger a random
number generator is utilised such that one out of every N events is
accepted.} for the 2012 data taking. 

The 2mu13 trigger requires two or more muon candidates, each of which passes
the single-muon trigger mu13.
The mu18\_mu8\_FS trigger requires at least one muon candidate which passes
the single-muon trigger mu18,
and subsequently employs the full-scan algorithm at the event-filter
to find two or more muon candidates with
$\pt > 18$ and $\pt > 8$\,GeV for leading and sub-leading muons, respectively.
The full-scan trigger processes the entire detector and utilises more
computing resources than the triggers which process only data in one
RoI. Computing resources, not bandwidth, is the limiting factor for
these triggers. The leading muon was required to have a {\pt} of at
least 18\,GeV in the full-scan dimuon triggers for this reason. 
The 3mu6 trigger requires three or more muon candidates, each of which passes
the single-muon trigger mu6.
The 2mu13, mu18\_mu8\_FS and 3mu6 triggers were used without prescale
for the 2012 data taking. The dimuon triggers used to select {\jpsi}
decays will be discussed in more detail in section 7. 

\subsection{Operation in the 2012 data taking}

\begin{figure}[tb]
\begin{center}
\subfigure[]
 {\includegraphics[width=0.35\textwidth]{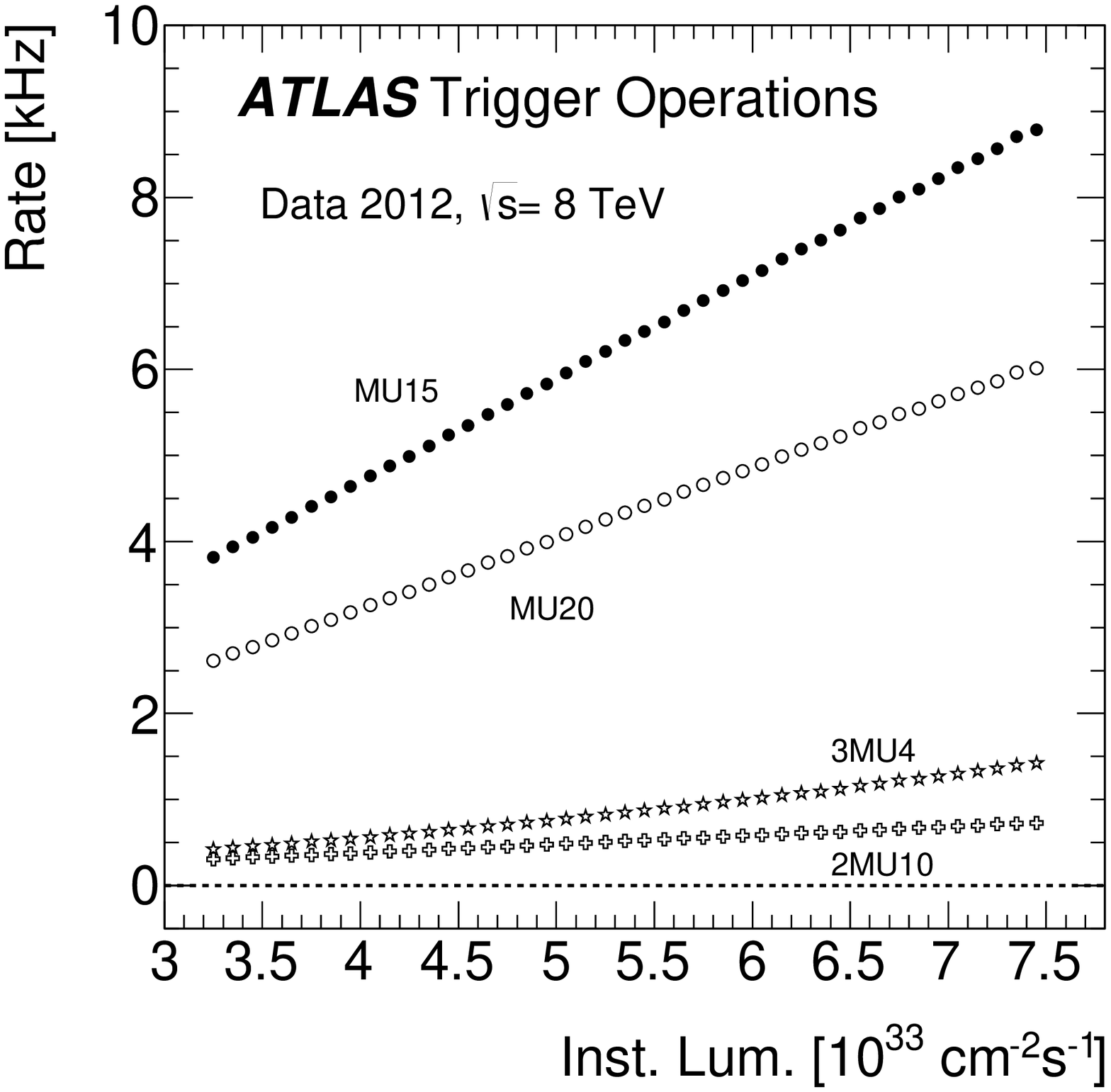}
  \label{fig:rates-l1}
}
\subfigure[]
 {\includegraphics[width=0.35\textwidth]{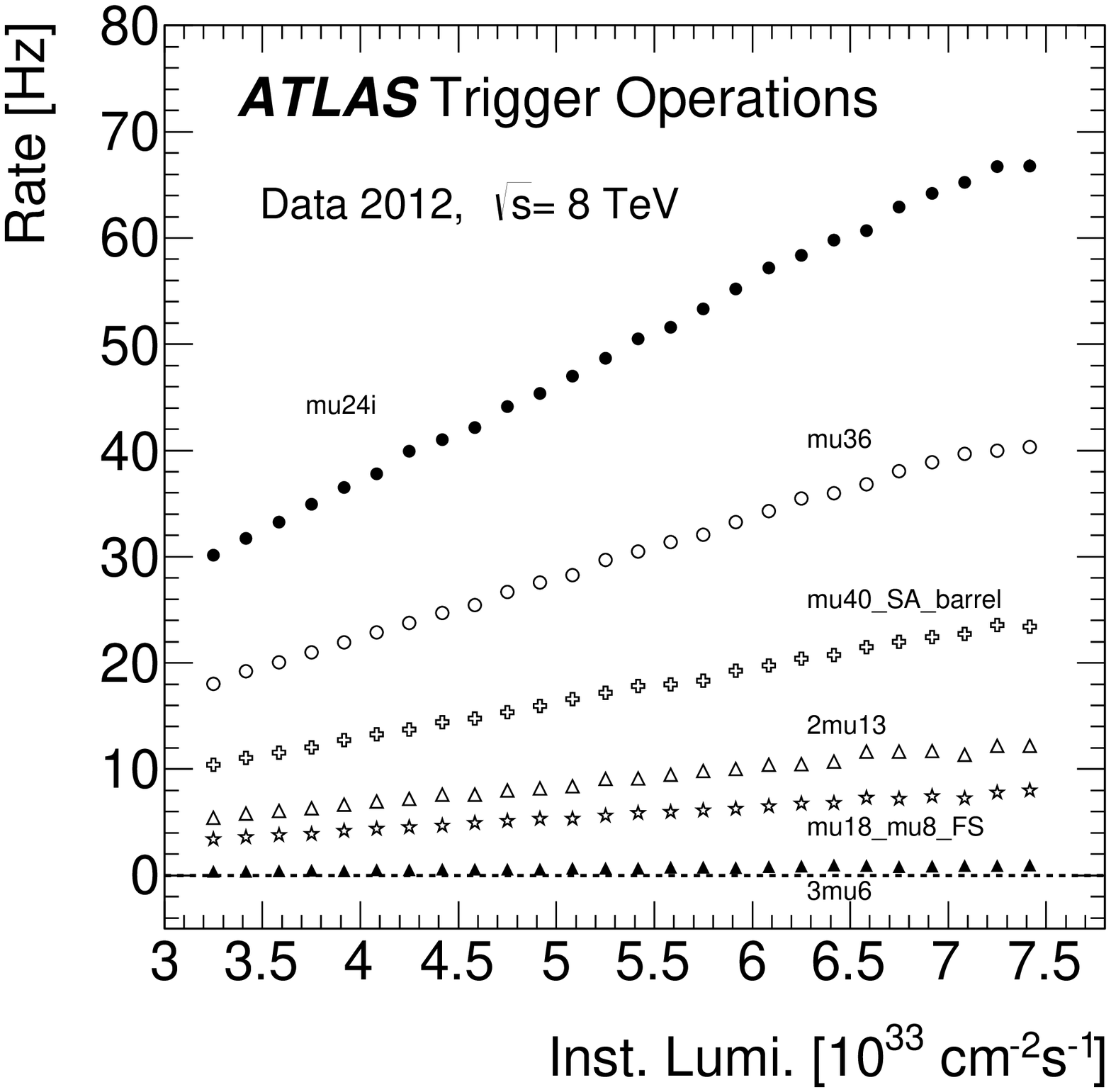}
  \label{fig:rates-ef}
  }
\end{center}
\caption{Trigger rates as a function of instantaneous luminosity 
(a) for selected muon triggers at Level-1 and (b) for  selected single- and multi-muon triggers at the event-filter
as denoted in the legend (see Table~\ref{tab:muon_chains} for details).}
\label{fig:rates}
\end{figure}

The typical maximum Level-1 rate was 70\,kHz.
The event acceptance was reduced at the event-filter which had an output
rate of 700\,Hz on average (with peaks of about 1\,kHz).
Of these rates, the single isolated muon trigger mu24i was 
fired at about 8.5\,kHz at Level-1 and at about 65\,Hz at the event-filter
for an instantaneous luminosity of 7$\times $10$^{33}$\,cm$^{-2}$s$^{-1}$.
Figure~\ref{fig:rates} shows the rates of
the single- and multi-muon triggers as a function of the instantaneous luminosity,
separately for the Level-1 and for the event-filter.
They are well described by a linear fit  with an approximately zero intercept.
This indicates a negligible contribution from effects 
not related to $pp$ collisions.
Typically the trigger rates were reduced by one to two orders of
magnitude at Level-2 and by a factor of a few at the event-filter for
the single and dimuon triggers. For example the rates were reduced by a factor of 28 at Level-2 (with respect to Level-1) and by a factor of 4.6 at the event-filter (with respect to Level-2)
for the mu24i trigger.
For the 2mu13 trigger, the rates were reduced by a factor of 71 at Level-2 and by a factor of 1.2 at the event-filter.


During data taking, the performance of the muon trigger was monitored in two stages. 
For quick online checks during data taking, 
the coverage in $\eta$--$\phi$ space and the distributions of some kinematic variables
were produced by the high level trigger algorithms.
A more detailed analysis was performed by calculating
efficiencies of triggers during the reconstruction stage of the  data processing.


%% file: dataset.tex
\section{Data samples and event selection}
\label{sect:sample-and-rec}

Several methods are used to measure the muon trigger
performance. This section describes the selection requirements
used to define the samples needed for the various methods.

\subsection{Methods to measure trigger performance}
\label{sect:method}

The tag-and-probe method relies on a pair of muons.
If one muon has caused the trigger to record the event (called the tag-muon),
the other muon serves as a probe (called the probe-muon) to measure the trigger
performance without any bias.
This method was applied to dimuon decays of  {\Zboson}-boson and
{\jpsi} meson candidates.
Alternatively, muons contained in events that were recorded by triggers other than the muon trigger can be used as an unbiased sample
to evaluate the efficiency of triggering on muons. 
This method was applied to events with muons from {\Wboson}-boson decays, either 
from top-quark or {\Wboson} + jets production.
A trigger based on the missing transverse momentum, as measured with the calorimeter,
was used to collect such samples.

Among these four samples,
the tag-and-probe method using {\Zboson} decays
provides the most precise determination of the efficiency
over a wide range of muon {\pt} ($10 \lesssim \pt \lesssim 100$\,GeV).
The tag-and-probe method using {\jpsi} decays provides
a coverage for lower {\pt} of the muon ($\pt \lesssim 10$\,GeV).
Muons from {\Zboson} decays are not frequently found to have {\pt}
$\gtrsim 100$\,GeV.  Events with muons from  top-quark and {\Wboson} + jets production
provide supplemental coverage at very high {\pt} 
($\pt \gtrsim 100$\,GeV). The muons from top-quark decays tend to have
a slightly larger {\pt} than those from the {\Zboson} decays due to the larger mass of the
top-quark. In the {\Wboson} + jet events, the {\Wboson} 
may recoil off of one or more high {\pt} jets. These higher {\pt}
{\Wboson}-bosons can then decay into muons with very high {\pt}. 
In addition, top-quark events and {\Wboson}+jet events offer 
important cross-checks in the overlapping $\pt$ region
that is also covered by the tag-and-probe method using $\Zboson$ decays.

\subsection{Data and Monte Carlo samples}
\label{sect:dataset}

Data were considered if recorded
under stable beam conditions and with all relevant sub-detector systems
fully operational. 

The trigger performance observed in the data is compared with the ATLAS Monte Carlo (MC)
simulation, which is the same as used for physics analysis. The
generated samples were then processed through a simulation of
the ATLAS detector based on 
\textsc{Geant4}~\cite{ref:geant4, ref:simulation}.
The environmental backgrounds due to radiation were not simulated.
The simulated events are overlaid with additional minimum-bias events
generated with \textsc{Pythia~8}~\cite{ref:pythia8} to account for the effect of pile-up interactions.

A sample of {\Zboson}-boson production was generated
using \textsc{Powheg-box}~\cite{ref:powheg} interfaced to \textsc{Pythia~8}~\cite{ref:pythia}.
A sample of the production of {\jpsi} mesons decaying to muon pairs was generated using \textsc{Pythia~8}, 
requiring at least two muons in the final state having
$\pt > 15$ and 2.5\,GeV.
Similarly to the {\Zboson}-boson production sample, a sample of top and antitop quark pair ({\ttbar}) events was generated using \textsc{Powheg-box}
interfaced to \textsc{Pythia~8}. Samples of single top-quark events were generated using \textsc{AcerMC}~\cite{ref:acermc}
interfaced to \textsc{Pythia~8} for the $t$-channel production, and 
using \textsc{Powheg-box} interfaced to \textsc{Pythia~8} for the $s$- and $Wt$-channel production. Samples of {\Wboson} boson production were generated using \textsc{Alpgen}~\cite{Mangano:2002ea}
interfaced to \textsc{Pythia~8}.
Samples of dijet events are used for background estimation,
and were generated using \textsc{Pythia~8}.


\subsection{Offline reconstruction}
\label{sect:offline_muon_rec}

The offline reconstructed muons are obtained by matching tracks found in the muon spectrometer
with those in the inner detector~\cite{ref:mcp_2012_paper}.
Muons are required to pass
various cuts to ensure a high quality inner detector track
and to be in a fiducial region of $\abseta<2.5$.
The muon momentum is calibrated by comparing the dimuon mass of {\Zboson} boson candidates
measured in data and MC~\cite{ref:mcp_2012_paper}.

The identification and reconstruction of the electrons, jets,
jets containing $B$-hadrons (called $b$-jets),
and missing transverse momentum ({\MET}) are necessary 
for the efficiency measurement with top-quark and {\Wboson}-boson candidates.

Electron candidates~\cite{ref:electron_id_2010,ref:electron_id_2011} are required to satisfy
$\et^{\mathrm{el}} > 25$\,GeV
and $|\eta^{\mathrm{el}}|<2.47$ excluding $1.37<|\eta^{\mathrm{el}}|<1.52$,
where $\et^{\mathrm{el}}$ is the transverse energy,
and $\eta^{\mathrm{el}}$ is the pseudorapidity of the electromagnetic cluster of energy deposits in the calorimeter.
Candidates are required to be isolated by means of
calorimeter- and track-based isolation parameters~\cite{ref:electron_isolation}.

Jets are reconstructed using the anti-$k_t$ jet clustering~\cite{ref:anti_kt}
algorithm with a radius parameter $R = 0.4$,
running on three-dimensional clusters of cells with significant calorimeter response~\cite{ref:topo_cluster}.
Their energies have object-based corrections applied as well as corrections for upstream material, non-instrumented material, and sampling fraction.
Jets are required to satisfy
$\ptjet > 25$\,GeV and $|\etajet| < 2.5$,
where $\ptjet$ is the transverse momentum, and
{\etajet} is the pseudo-rapidity of the jet.
Duplication between electron and jet objects is avoided
by removing the jet closest to an electron
if their separation is $\Delta R < 0.2$.

The $b$-jets are identified among the reconstructed jets
with an artificial neural network using variables
that exploit the impact parameter, the secondary vertex and the
topology of $b$- and $c$-hadron
weak decays~\cite{ref:bjet_calibration}.
An identification criterion with 70\,\% efficiency is chosen,
as evaluated on jets in a simulated $\ttbar$ sample with $\pt > 20 \GeV$ and $\abseta < 2.5$.

Hadronically decaying taus are reconstructed using clusters in the electromagnetic and hadronic calorimeter
~\cite{ref:detector_paper}. A Boosted Decision Tree tau
identification method is used to select candiates with a 55-60 \%
efficiency. Tau candidates are required to have a charge $\pm 1$ and
only one or three tracks in a cone of radius $\Delta R < 0.2$. 

Photons are identified by  electromagnetic cluster of energy deposits
in the calorimeter similar to electron identification ~\cite{ref:photon_id}. In the case of
photons, isolated electromagnetic clusters without matching tracks are
classified as unconverted photon candiates. Clusters matched to a pair
of tracks that are consistent with the hypothesis of a $\gamma
\rightarrow e^+e^-$ conversion process are classified as converted
photon candiates. 

The {\MET} is calculated using the reconstructed jets,
electrons, muons, $\tau$ leptons, photons, as well as calorimeter energy
clusters not associated with these physics objects~\cite{ref:etmiss}.

In this paper,  reconstructed objects (using
algorithms applied after the event is recorded) are distinguished from trigger objects (formed either at Level-1, Level-2, or the event-filter
during the fast online reconstruction of the event). 

\subsection{Event selection for the {\Zboson}-boson sample}
\label{sect:dataset_zselection}
For the selection of the {\Zboson}-boson sample,
events are required to pass either the isolated single-muon trigger mu24i
or the single-muon trigger mu36.

A pair of oppositely charged muons with invariant mass, $m_{\mu \mu}$,
consistent with the mass of the {\Zboson} boson,
$|m_Z - m_{\mu \mu}| < 10$\,GeV, is required.
The two muons are required to originate from the
same interaction vertex.
If one of the two muons has $\pt > 25$\,GeV and
is isolated, $\Sigma_{\Delta R < 0.2} \pt^{\mathrm{trk}} / \pt < 0.1$,
it is a candidate for the tag-muon, and the other muon is a candidate
for the corresponding probe-muon.
From a pair of muons, two candidate tag- and probe-muons
are allowed.
Furthermore, the tag-muon candidate must have an angular distance of $\Delta R < 0.1$ to
an event-filter combined-muon that passes either the mu24i or mu36 trigger.
In addition, the probe-muon candidate has to be isolated,
$\Sigma_{\Delta R < 0.2} \pt^{\mathrm{trk}} / \pt < 0.1$.


The probe-muon is matched to a trigger object if it lies within a distance
$\Delta R < 0.1$ from an event-filter combined-muon 
and $\Delta R < 0.5$ from a Level-1 trigger object.
The trigger efficiency is defined as the fraction of probe-muons
that are associated with at least one trigger muon-object after applying the above
criteria. 

\subsection{Event selection for the {\jpsi} meson sample}
\label{sect:jpsi_tap_selection}

Due to rate restrictions, samples of {\jpsi} candidates were selected using
the two dedicated triggers as in Table~\ref{tab:muon_chains}.
One trigger requires
a pair of muons found by the event-filter full-scan with a mass consistent with
that of the {\jpsi}, with at least one muon with $\pt > 18$\,GeV.
It is used to determine the efficiency at Level-1 and Level-2.
The other trigger requires a pair of muons found by Level-1 and Level-2 with
the same requirements as above.
It is used to determine the efficiency at the event level with respect to the Level-1 and Level-2.
Then the total efficiency can be obtained by multiplying these two partial efficiencies.

All combinations of oppositely charged offline muons are considered as
{\jpsi} candidates if each of the muon tracks satisfies
$|d_0| < 0.2$\,~mm,
where $d_0$ is the distance of closest approach between the inner detector track
and the proton-proton interaction in the plane transverse to the beam.
The two inner detector tracks that are associated with the two muon tracks are
refitted under the assumption that they originate from the same vertex.
The invariant mass constructed from the refitted tracks, $m_{\mu\mu}$,
is required to be consistent with the
 {\jpsi} mass, $|m_{\jpsi} - m_{\mu\mu}| < 0.3$\,GeV.
To enhance the fraction of muons originating from a {\jpsi} decay a further requirement is
made on  $L_{xy}$ , the signed two-dimensional decay length of the {\jpsi}.
The variable $L_{xy}$ is defined as $L_{xy} \equiv \vec{L} \cdot \vec{\pt^{\jpsi}}
 / \pt^{\jpsi}$ with $\vec{L}$ being the vector originating from the
 proton-proton interaction vertex. A requirement of  $L_{xy} < 1$~mm is made on the muons.
%
The requirements on $d_0$ and $L_{xy}$ are used to suppress non-prompt muons,
such as those from  the decays of $B$-hadrons~\cite{ref:jpsi_paper}.

The fact that these two dedicated triggers were used to select {\jpsi} candidates
implies that the {\jpsi} mesons are boosted and therefore the spacial distance
between the two muons from the decays is small.
To ensure correct one-to-one matching between trigger and offline muons,
the distance between them is gauged by
the separation of the impact points of the tracks at the
locations of the RPC and TGC detectors after extrapolation based on the refitted inner detector track parameters.
If one of the two muons has $\pt > 18$\,GeV and its distance from an event-filter combined-muon that passes the mu18 trigger is within $\Delta R < 0.08$, as evaluated by using the extrapolated
positions, it is considered as a tag-muon.
If the other muon is beyond the distance of $\Delta R > 0.2$ from the
tag-muon, at  the extrapolated positions, it is regarded as a probe-muon.
The $\Delta R$ cut value is sufficiently large compared to the typical dimensions
of the Level-1 trigger segmentation, as described in Sect.~\ref{sect:trigger_l1}.
A probe-muon is matched to trigger objects, if it is within $\Delta R < 0.12$ from a Level-1 muon object and an event-filter combined-muon. 

\subsection{Selection of top quark and {\Wboson} + jets candidate events}

The  top quark and {\Wboson} + jets candidate events have to pass a trigger that requires
$\MET (\mathrm{calo}) > 80$\,GeV,
where $\MET (\mathrm{calo})$ is the magnitude of the
missing transverse momentum as measured using only the calorimeter information.
Several additional cuts are then imposed to 
remove events with noise bursts in the calorimeters
and those with cosmic-ray showers.

The muon candidate is required to have
$\pt > 40$\,GeV and 
$|z_0| < 2$\,mm,
where $z_0$ is the track impact parameter in the $z$-direction with
respect to the proton-proton interaction vertex.
The probe-muon is required to be isolated from neighbouring jets and energy depositions in the
calorimeter. Probe-muons are required to satisfy $\Sigma_{\Delta R <
  0.3} p^{\rm trk}_T/\pt < 0.05$
and $\Delta R_{\min} (\mathrm{jet,muon}) > 0.4$, where $\Delta
R_{\min} (\mathrm{jet,muon})$ is the minimum distance between the muon
and any jet. In addition, no other muon with $\pt > 25$\,GeV is allowed.

Events are further required to have 
$\MET > 20$\,GeV and
$m_{\mathrm{T}}^{\Wboson} + \MET > 60$\,GeV,
where $m_{\mathrm{T}}^{\Wboson}$ is the transverse
mass\footnote{Transverse mass is defined as $m_{T}^{2} = m^{2} +
  p_{x}^{2} + p_{y}^{2}$  and has the useful propriety that
it is invariant under Lorentz boosts along the beam direction.} of the
{\Wboson} candidate.
The {\Wboson} is reconstructed with four-vectors of the {\MET} and the muon.

For the top quark sample, there must be at least three jets
with at least one $b$-jet.
For the {\Wboson} sample, there must be one or two jets
with no $b$-jets. Events with an electron are rejected.


%% file: purity.tex
\section{Trigger purity}
\label{sect:purity}

The trigger purity is defined as the fraction of muon triggers that can be associated 
to an offline muon.
The $\Delta R$ distance between the trigger object and the offline muon
was used to define this matching.

\begin{figure}[htb]
\begin{center}
\subfigure[]
  {\includegraphics[width=0.32\textwidth]{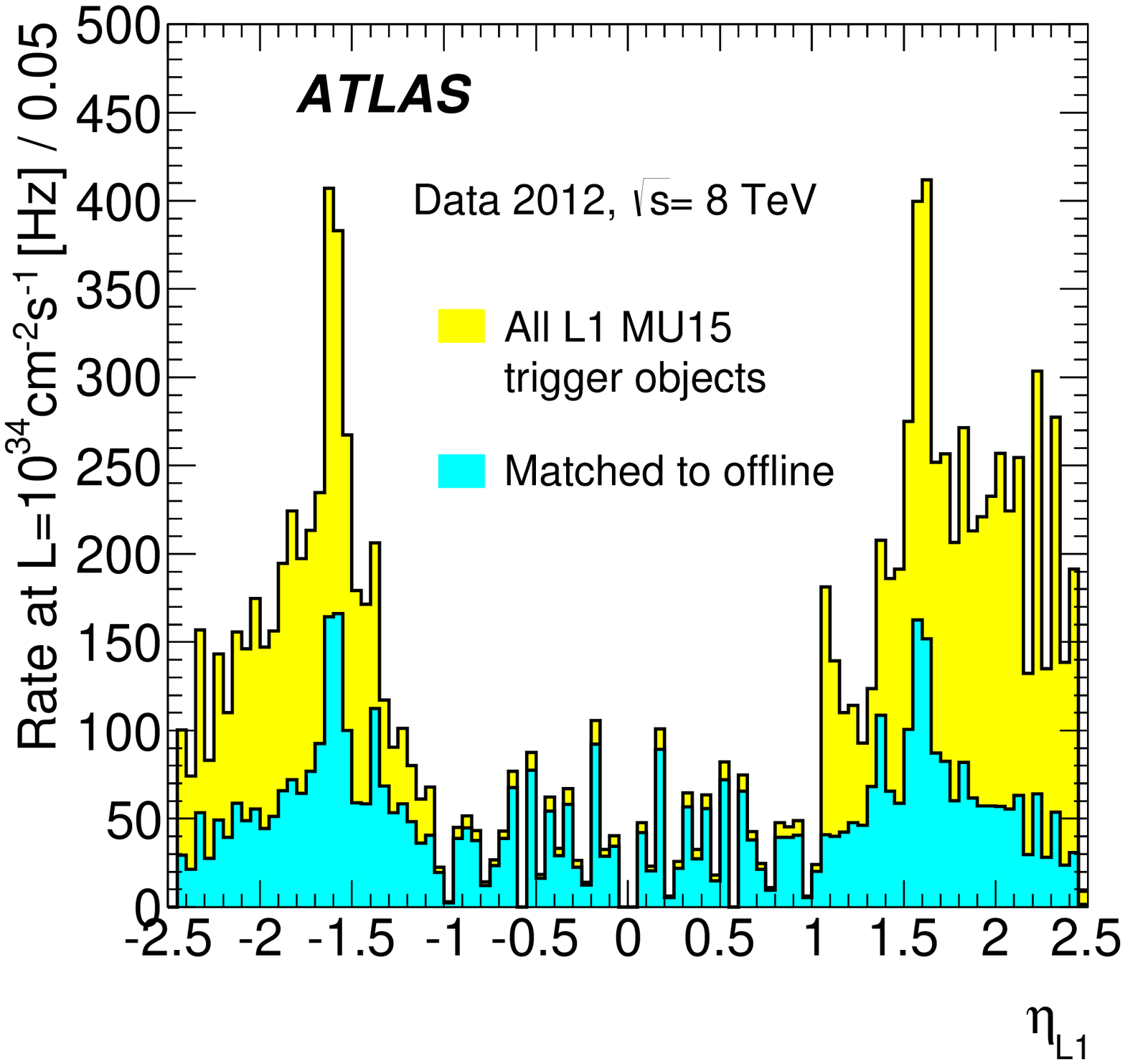}
  \label{fig:l1_purity_eta}
  }
\subfigure[]
  {\includegraphics[width=0.32\textwidth]{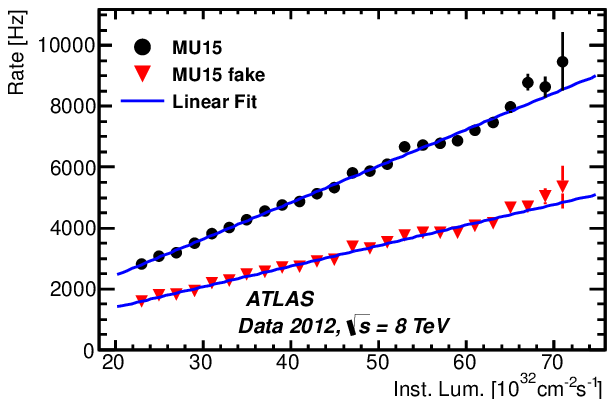}
   \label{fig:l1_purity_rate}
  }
\end{center}
\caption{Trigger  rate of the Level-1 MU15 as a function of
  (a)  pseudorapidity $\eta_\mathrm{L_1}$ of all the trigger objects (light histogram) and of the ones 
  associated with offline reconstructed muons (dark histogram) and
  (b)  instantaneous luminosity,  for all triggers (dots)  and for the fake ones not-associated with offline-reconstructed
   muons (triangles) with the lines representing the results of the corresponding linear fits. } 
\label{fig:l1_purity}
\end{figure}

The  $\eta$ distribution of the Level-1 MU15 object that seeds
the mu24i event-filter is shown in
Fig.~\ref{fig:l1_purity_eta} for all triggers and for those associated
with a reconstructed offline muon. 
\begin{figure}[htb]
\begin{center}
\subfigure[]
  {\includegraphics[width=0.32\textwidth]{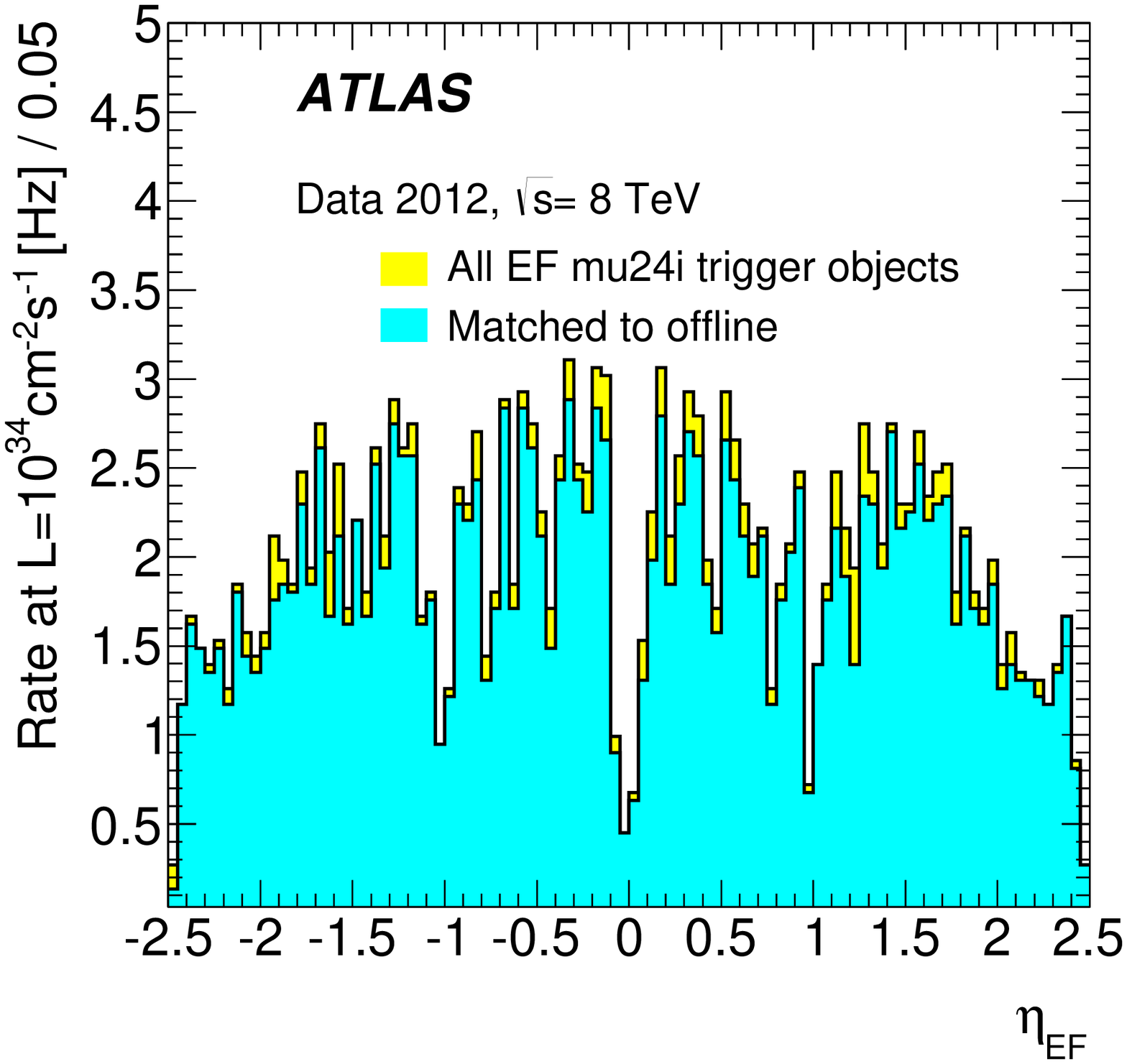}
  \label{fig:ef_purity_eta}
  }
\subfigure[ ]
  {\includegraphics[width=0.32\textwidth]{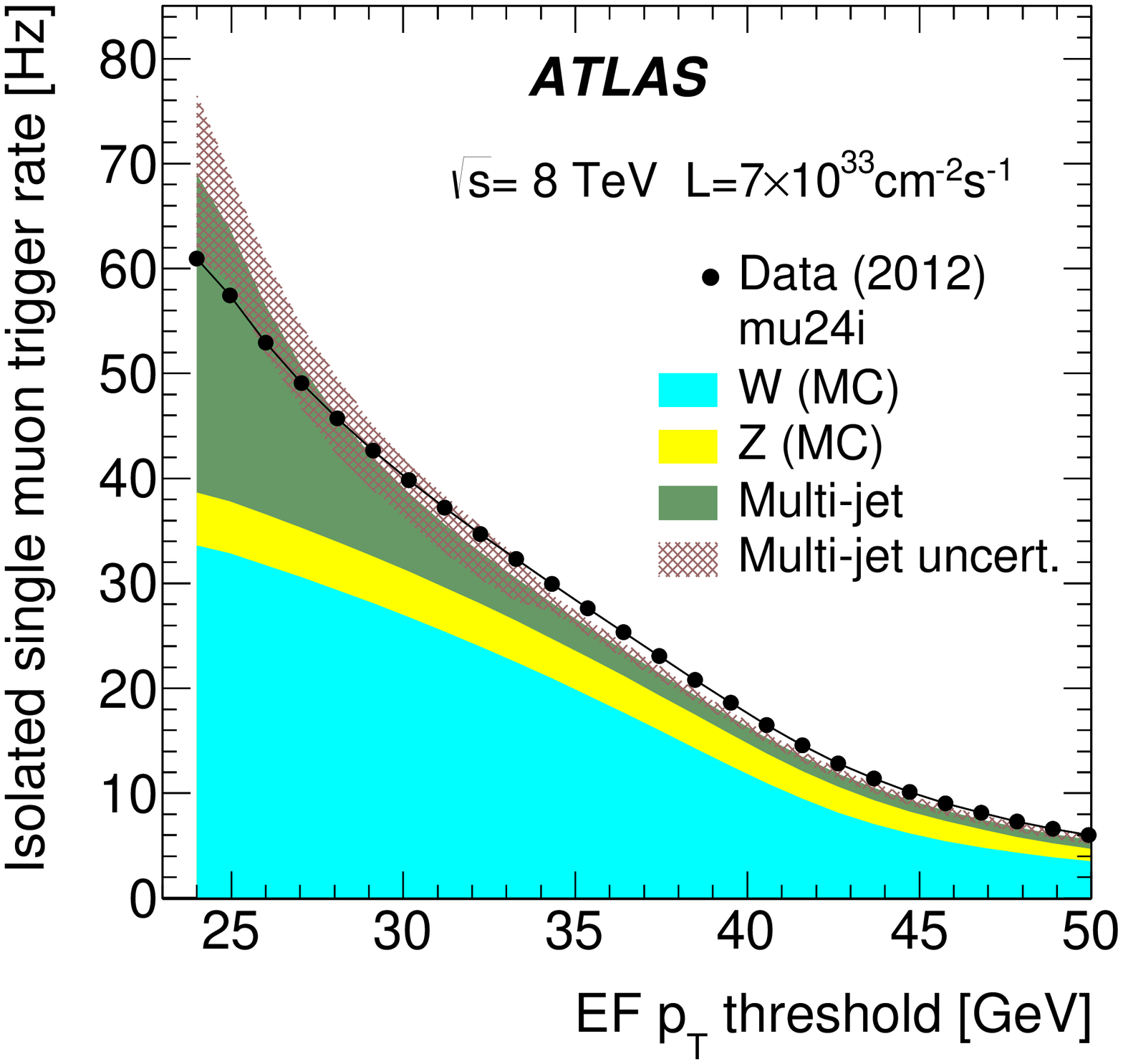}
  \label{fig:ef_purity_physcont}
  }
\end{center}
\caption{Rate of the isolated single-muon trigger, mu24i, at
  the event-filter 
  (a)  as a function of pseudorapidity $\eta_\mathrm{EF}$ for all combined-muons (light histogram) and for the ones 
  associated with offline reconstructed muons (dark histogram);
  (b)  as a function of the transverse momentum {\pt} threshold at the event-filter (EF) at an instantaneous luminosity of 7$\times$10$^{33}$\,cm$^{-2}$s$^{-1}$,
   for  combined-muons in the data (dots) compared to the expectations from {\Wboson}- and  {\Zboson}- bosons production and from 
the data-driven estimate for multi-jet production, as described in the legend.   }
\label{fig:ef_purity}
\end{figure}
%
No explicit cut on offline muon {\pt} was applied in the association
between trigger and offline objects.
Figure~\ref{fig:l1_purity}  shows that the Level-1 rate is dominated by triggers without associated 
offline muons (called fake triggers).
The overall trigger purity (fraction of Level-1  rate from true
muons ) is 40\,\%. Most of the  Level-1 fakes originates
in the end-cap. 
The cause of these fakes in the endcap region was extensively investigated~\cite{ref:nsw_tdr}, and is understood as mainly due to charged
particles, for instance protons, produced in large amounts of dense
material such as the  toroid coils and shields.
Figure~\ref{fig:l1_purity_rate} shows the MU15 trigger rate as a function of
the instantaneous luminosity. Also shown is 
the rate due to fake triggers.
The error bars show statistical uncertainties only.
Both the total rate and the fake rate at Level-1
scale linearly with the instantaneous luminosity.

%
Figure~\ref{fig:ef_purity_eta} shows
the $\eta$ distribution of the trigger objects recorded with the isolated single-muon trigger at
the event-filter. The fake triggers, not associated to an offline reconstructed muon, are rejected by the subsequent High-Level-Trigger decisions,
and a purity of about 90\,\% is achieved.
The physics origin of muons at the event-filter is illustrated in Fig.~\ref{fig:ef_purity_physcont},
which shows the expected composition of the trigger rate of the
isolated single-muon as a function of the lower threshold value on the
muon {\pt}.
The vertical scale gives the trigger rate as a function of  {\pt} at an instantaneous luminosity of 7$ \times$10$^{33}$\,cm$^{-2}$s$^{-1}$.
The expectations for {\Wboson} and {\Zboson} production were evaluated
by using MC simulations with their predicted cross sections. Multi-jet
production, where one or more jets produce a muon from the decay of a
heavy quark or from 
a pion or kaon decay in flight, also contribute to this rate. The multi-jet contribution was evaluated in a data-driven
approach as described below.

A multi-jet enriched control-region  is obtained by using
events that are triggered by a single-muon trigger with the
same {\pt} threshold but without isolation requirement.\footnote{This trigger was active but with a prescale factor of 10.}
The control-region is defined by inverting the trigger isolation criteria,
by requiring at least one jet in an event, and by requiring matching to an offline muon
to remove the fake contribution.
The multi-jet contribution in the signal region is estimated by the
following procedure. The fraction of multi-jet events in the signal region is
taken from dijet MC simulation. The total normalization for the
multi-jet contribution is then evaluated in the control-region. The
contribution to the signal region is then taken as the total estimated
multi-jet contribution  weighed by the signal fraction from simulation.
The uncertainty of this estimation is dominated by the statistical uncertainty
in the control-region/signal-region  transfer factors from MC simulation, and is shown in Fig.~\ref{fig:ef_purity_physcont}.
The rate was evaluated as a function of the {\pt} threshold on the event-filter combined-muon.
As shown in Fig.~\ref{fig:ef_purity_physcont}, at $\pt = 24$\,GeV
about 60\,\% of the events triggered by mu24i are due to muons from {\Wboson} and {\Zboson} production.


%% file: resolution.tex
\section{Resolution}
\label{sect:resolution}

The tag-and-probe method applied to {\Zboson}-boson  candidates was used to evaluate
the quality of the $\pt$, $\eta$ and $\phi$ determination at the event-filter,
compared to the offline reconstruction.
The online algorithms are nearly identical to the offline versions but have some
simplifications in the pattern recognition because of timing
constraints. Additionally, the offline reconstruction uses updated
calibration and alignment corrections not available at the time the
data was recorded. 
Therefore, finite difference can be expected
even when the event-filter combined muon is compared with
the offline muon that is also reconstructed by combining
the inner detector and muon detectors.

\begin{figure}
\begin{center}
\includegraphics[width=0.32\textwidth]{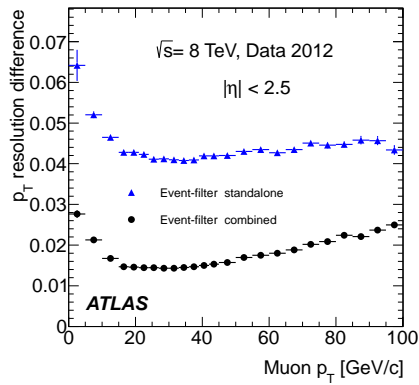}
\label{fig:ef_pt_resolutions_barrel}
\end{center}
\caption{Resolution difference in transverse momentum {\pt} determination in the offline and in the event-filter reconstruction, as a function of
   {\pt} of the  offline muon. }
\label{fig:ef_pt_resolution}
\end{figure}

The offline momentum resolution is $< 3.5$ \%  up to transverse momenta
 $p_{T}$  of 200 GeV and $<$ 10\% up to 1 TeV
~\cite{ref:mcp_2010_paper}.
The residual of the trigger-reconstructed {\pt} with respect
to the offline value is defined as
$\delta_{\pt} = \frac{1/p^{\mathrm{trigger}}_T - 1/\pt}{1/\pt}$,
where $p^{\mathrm{trigger}}_{T}$ is the
transverse momentum reconstructed by the trigger,
and {the \pt} is that of the offline muon. 
The resolution difference between the trigger and offline reconstruction
was defined as the standard deviation of a Gaussian
function fitted to the $\delta_{\pt}$ distribution.
Figure~\ref{fig:ef_pt_resolution} shows the {\pt} resolution
differences, as a function of the offline muon {\pt}, of the event-filter stand-alone
and event-filter combined muons in the barrel and endcap regions.
The {\pt} resolution difference is about 2\,\% and 5\,\% for event-filter combined and event-filter stand-alone
muon, respectively.

\begin{figure}[tb]
\begin{center}
\subfigure[]
  {\includegraphics[width=0.32\textwidth]{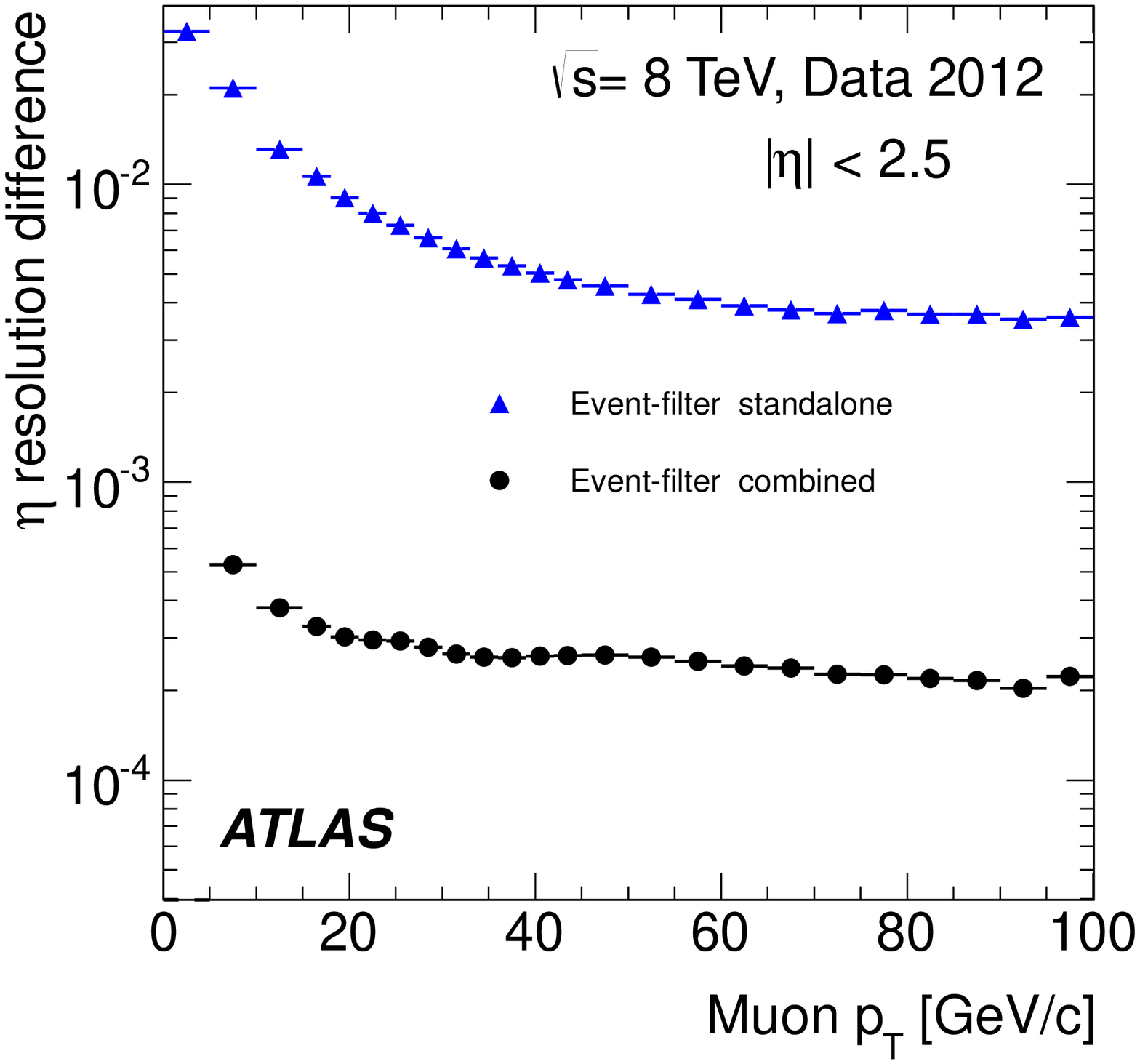}
  \label{fig:ef_resolutions_eta}
  }
\subfigure[]
  {\includegraphics[width=0.32\textwidth]{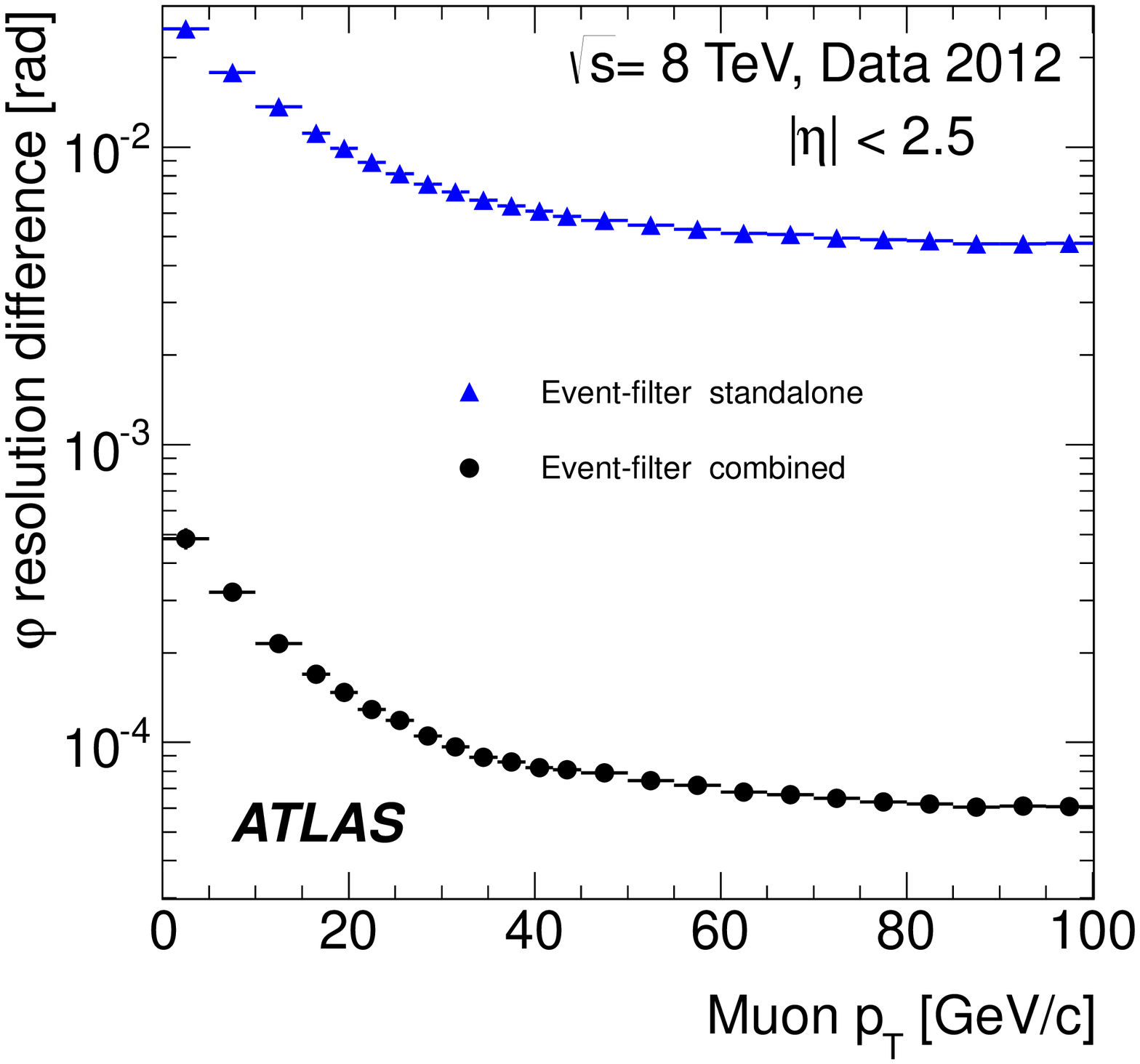}
  \label{fig:ef_resolution_phi}
  }
\end{center}
\caption{
  Resolution difference in the (a) pseudorapidity $\eta$ and (b)~ azimuthal angle $\phi$ determination
  in the offline and in the event-filter reconstruction, as a function of
  {\pt} of the offline muon.}
\label{fig:ef_resolution}
\end{figure}

The resolution differences of the $\eta$ and $\phi$ determination
were examined similarly by defining the residual as the absolute value of the difference
between the trigger and offline reconstructed values.
Figure~\ref{fig:ef_resolution} shows the $\eta$ and $\phi$ resolution
differences of the event-filter muons.
%
It shows that the
trigger--offline matching criterion used in the efficiency measurements, for instance
$\Delta R < 0.1$ for the tag-and-probe method using {\Zboson } bosons (see Sect.~\ref{sect:dataset_zselection}),
is sufficiently loose.


%% file: ztap.tex
\section{Efficiency measurements with {\Zboson} boson candidates  } 
\label{sect:ztap_efficiency}

In the next several sections, measurements of the efficiency
of the muon trigger in different kinematic regions are presented, preceeded by a discussion of systematic uncertainties. 
The efficiency is primarily measured as a function of muon $\pt$.
In addition, the efficiency is measured in two-dimensions,
for instance in $\eta$ and $\phi$ bins, and compared to
the simulated one.
To more accurately model data, all ATLAS physics analysis which use
events selected with the muon trigger are provided with the ratios of measured to simulated efficiencies to
make small corrections to the simulated samples.




\subsection{Systematic uncertainty}
\label{sect:ztap_syst}

In the following, sources of systematic uncertainty are discussed and 
the  quoted uncertainty values are presented for the
efficiency measured in the region of $25 < \pt < 100$\,GeV.

\begin{figure}[htb]
\begin{center}
\subfigure[]
  {\includegraphics[width=0.38\textwidth]{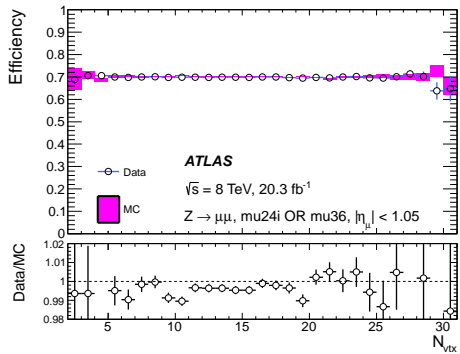}
  \label{fig:ztap_pileup_barrel}
  }
\subfigure[]
  {\includegraphics[width=0.38\textwidth]{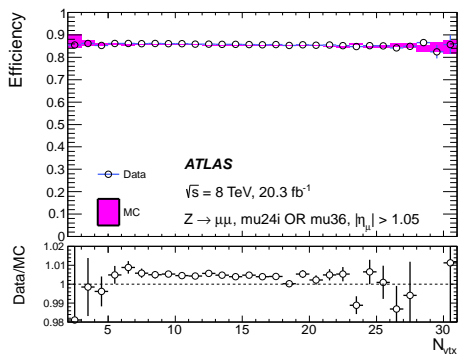}
  \label{fig:ztap_pileup_endcap}
  }
\end{center}
\caption{Efficiency to pass either mu24i or mu36 triggers,
  as a function of the number of reconstructed
  vertices in an event,
  N$_{\mathrm{vtx}}$ in 
  (a) the barrel region, and in (b) the endcap region, for data (dots) and MC simulation (bands).
  The lower panels show the ratio of the
  efficiencies in data and in MC simulation.
  The error bars reflect statistical uncertainties only.
}
\label{fig:ztap_pileup}
\end{figure}

\begin{itemize}
\item Dependence on pile-up interactions:\\
the efficiency was measured as a function of
the number of reconstructed vertices, Nvtx, separately for data and MC simulation,
as shown in Fig.~\ref{fig:ztap_pileup}.
The efficiency is largely independent of the number of pile-up interactions.
Separate linear fits to the data and MC simulation were performed in the range from Nvtx=5 to Nvtx=30 and extrapolated out to Nvtx=50.  The dependence on the fit range was observed to be negligible.  The largest difference observed between the fits in data and MC simulation were observed to be 0.1 (0.5) \% in the barrel (endcap). This difference is taken as an estimate of the systematic uncertainty due to the presence of pile-up interactions.

\item Correlation between tag- and probe-muons from {\Zboson} decays:\\
for medium {\pt}, tag- and probe-muons
tend to be back-to-back in $\phi$.
Since the barrel and endcap have 16-fold and 12-fold symmetries, respectively,
this can potentially lead to some bias;
a tag-muon from a {\Zboson} -boson decay inside a highly efficient region of
the detector tends to be accompanied
by a probe-muon in a region of high efficiency.
This effect is evaluated by adding a requirement
to the tag and probe pairs to prevent them from being back-to-back,
$\Delta\phi\mathrm{(tag, probe)} < \pi - $\,0.1,
where $\Delta\phi\mathrm{(tag, probe)}$ denotes the azimuthal angle between the
tag- and probe-muons. The resulting uncertainty in the efficiency determination
is 0.3\% (0.2\%) in the barrel (endcap) region.

\item Matching between probe-muon and trigger muon:\\
this effect was estimated by changing the $\Delta R$ 
thresholds of the matching criteria. The change in the efficiency determination
was found to be negligible.

\item Probe-muon momentum scale and resolution:\\
this effect was estimated by changing the momentum scale and momentum resolution 
for the  probe-muon by their respective uncertainties, as determined from the
calibration using {\Zboson}-bosons.
The resulting change in efficiency was negligible. 

\item Probe-muon selection criteria:\\
this effect was estimated by changing, typically by 10\,\%, the cuts
in various selection criteria, leading to negligible changes in the efficiency determination.

\item Background contribution:\\
the amount of background was estimated by using the dijet,
{\ttbar}, and {\Wboson} MC simulations and the effect on the efficiency determination was found to be negligible~\cite{ref:mtsg_2011confnote}.
Also, varying the {\Zboson} mass window cut gave negligible effect.

\item  MC modelling:\\
the sensitivity of the efficiency determination to the MC modelling was tested
by comparing  samples generated with a different
MC generator, namely by adding \textsc{Sherpa}~\cite{ref:sherpa}.
Again, the change in efficiency  was found to be negligible~\cite{ref:mtsg_2011confnote}.

\item  Dependence on {\pt}:\\
  after correcting the MC efficiency in $\eta$ and $\phi$ so as to reproduce the one observed in the data ,
  any residual deviations between data and MC in the $\pt$ dependence are taken as systematic uncertainty.
  This resulted in a 0.4\,\% effect.

\item Probe-muon charge dependence:\\
  it was estimated by comparing the efficiencies measured with positively charged and negatively charged probe-muons.
  The estimated uncertainty is 0.2\,\% in the endcap region.
\end{itemize}

The individual systematic uncertainties are added in quadrature to obtain
the total systematic uncertainty, resulting in 
0.6\,\% for the efficiency measured in the region of $25 < \pt < 100$\,GeV.

\subsection{Single-muon triggers: mu24i, mu36}
 
\begin{figure}[htb]
\begin{center}
\subfigure[]
  {\includegraphics[width=0.38\textwidth]{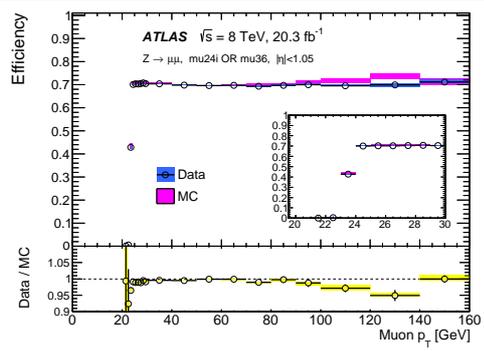}
  \label{fig:ztap_mu24i_pt_barrel}
  }
\subfigure[]
  {\includegraphics[width=0.38\textwidth]{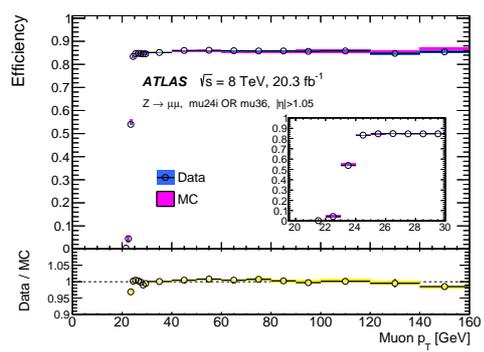}
  \label{fig:ztap_mu24i_pt_endcap}
  }
\end{center}
\caption{Efficiency of passing either the mu24i or mu36 trigger as a function of the probe-muon transverse momentum {\pt},
   for
  (a) the barrel region and (b) the endcap region, for data (dots) and MC simulation (bands).
  The lower panels show the ratio of the data and MC efficiencies. The error bars include both statistical and
  systematic uncertainties.}
\label{fig:ztap_mu24i_pt}
\end{figure}

Requiring events to pass either 
the mu24i or the mu36 trigger
serves as a general-purpose single-muon triggers for many physics analyses.
Figure~\ref{fig:ztap_mu24i_pt} shows the efficiency
to pass either the mu24i or the mu36 trigger
as determined in the barrel and endcap regions.
The efficiency was measured as a function of the {\pt} of the reconstructed probe-muon for both data and simulation. The efficiency in the
simulation is seen to match that of the data over a wide {\pt} range. 
The slight excess in simulation in the {\pt} bin
centred at 130 GeV was studied in detail. High {\pt} muons from {\Zboson}-boson decays tend to be slightly more forward where there is the
largest difference in trigger efficiency between data and simulation.

The efficiency curve turns on sharply around the threshold,
reaching a plateau already around $\pt \sim 25$\,GeV.
In order to quantitatively evaluate the turn-on behaviour and the agreement between
data and MC simulation, a fit was made using a Fermi function $f(p_{T})$.\footnote{
The functional form is $\frac{a}{1+\exp{\{b(c-\pt)\}}}$,
where $a$ indicates the plateau value, $b$ the steepness of the 
turn-on slope, and $c$  the threshold value.}
From the fit, the low edge of the efficiency plateau region was defined
as the value of {\pt} for which the efficiency decreases by 1\,\% from the plateau value.
Table~\ref{tab:ztap_mu24i_turnon_threshold} shows these evaluated
plateau values as well as the location of the low edges of the plateaus.
\begin{table*}
\caption{Result of fitting a Fermi function to the efficiency turn-on
  curve as a function of transverse momentum {\pt}
  for the single-muon trigger, for data and MC simulation.
  The location  in {\pt} of the low edge of the plateau region is
  defined such that the efficiency decreases by 1\,\% from the plateau
  value.}
\label{tab:ztap_mu24i_turnon_threshold}
\begin{center}
\begin{tabular*}{\textwidth}{@{\extracolsep{\fill}}llrrrr@{}}
\hline
                          &               & \multicolumn{2}{c}{Data}   & \multicolumn{2}{c}{MC} \\
Trigger                   &               & Plateau value & Low edge & Plateau value & Low edge \\
\hline
Either mu24i or mu36 & Barrel & 70.1\,\%        &  24.3\,GeV  & 70.3\,\%        & 24.0\,GeV  \\
                          & Endcap        & 85.6\,\%        &  24.8\,GeV  & 85.3\,\%        & 24.7\,GeV  \\ \hline
\end{tabular*}
\end{center}
\end{table*}
The single-muon trigger that requires either the mu24i or mu36 trigger
exhibits a plateau efficiency for physics analysis with muon $\pt >
25$\,GeV. The efficiency plateau is smooth at $\pt=36$\,GeV
indicating that there is no inefficiency due to the isolation
requirement in this sample.

\begin{figure}
\begin{center}
\subfigure[]
  {\includegraphics[width=0.38\textwidth]{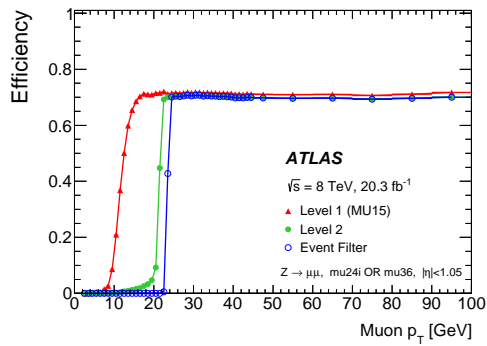}
  \label{fig:ztap_mu24i_pt_l1l2ef_barrel}
  }
\subfigure[]
  {\includegraphics[width=0.38\textwidth]{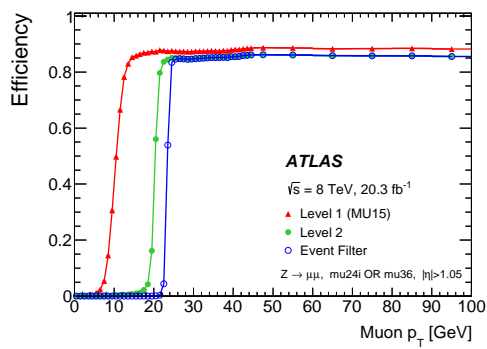}
  \label{fig:ztap_mu24i_pt_l1l2ef_endcap}
  }
\end{center}
\caption{Efficiency of passing either the mu24i or mu36 trigger as functions of the probe-muon transverse momentum {\pt},
   for the three trigger levels, Level-1, Level-2 and event-filter,  in the data for
  (a) the  barrel region and (b) the endcap region.
  The error bars show the statistical uncertainties only.}
\label{fig:ztap_mu24i_pt_l1l2ef}
\end{figure}

Figure~\ref{fig:ztap_mu24i_pt_l1l2ef} shows the efficiency
of requiring to pass either mu24i or mu36 triggers, as measured
separately for the three trigger levels, Level-1, Level-2 and event-filter.
The trigger selection becomes tighter and the efficiency turn-on
becomes sharper as the trigger level increases. The plateau efficiency is mostly determined by Level-1.
The high level trigger efficiency with respect to Level-1 is about 98\,--\,99\,\%.

\begin{figure}[tb]
\begin{center}
\subfigure[]
  {\includegraphics[width=0.35\textwidth]{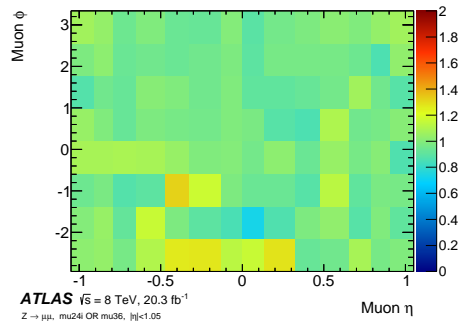}
  \label{fig:ztap_mu24i_etaPhi_barrel}
  }
\subfigure[]
  {\includegraphics[width=0.35\textwidth]{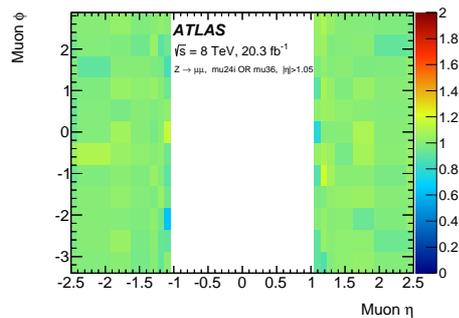}
  \label{fig:ztap_mu24i_etaPhi_endcap}
  }
\end{center}
\caption{Ratio of the data and MC efficiencies to pass either the mu24i or the mu36 trigger, 
   in bins of the probe-muon $\eta$ and $\phi$ in (a) the barrel region and (b) the endcap region.}
\label{fig:ztap_mu24i_etaPhi}
\end{figure}

Figure~\ref{fig:ztap_mu24i_etaPhi} shows the ratio of 
the data and MC efficiencies to pass either the mu24i or the mu36 trigger, as determined in bins of
$\eta$ and $\phi$ of the probe-muon,
for the barrel and endcap regions.
The measurement was performed for muons with $\pt > 25$\,GeV.
The bins in $\eta$ and $\phi$ are fine enough to
reflect the hardware segmentation of the Level-1 detectors but coarse
enough to have sufficient statistics in each bin.
The typical size of the statistical uncertainty is less than 1\,\%, except for a few specific areas
where the uncertainty is about 3\,\%.

\subsection{Other single-muon triggers}

\begin{figure}[tb]
\begin{center}
\subfigure[]
  {\includegraphics[width=0.38\textwidth]{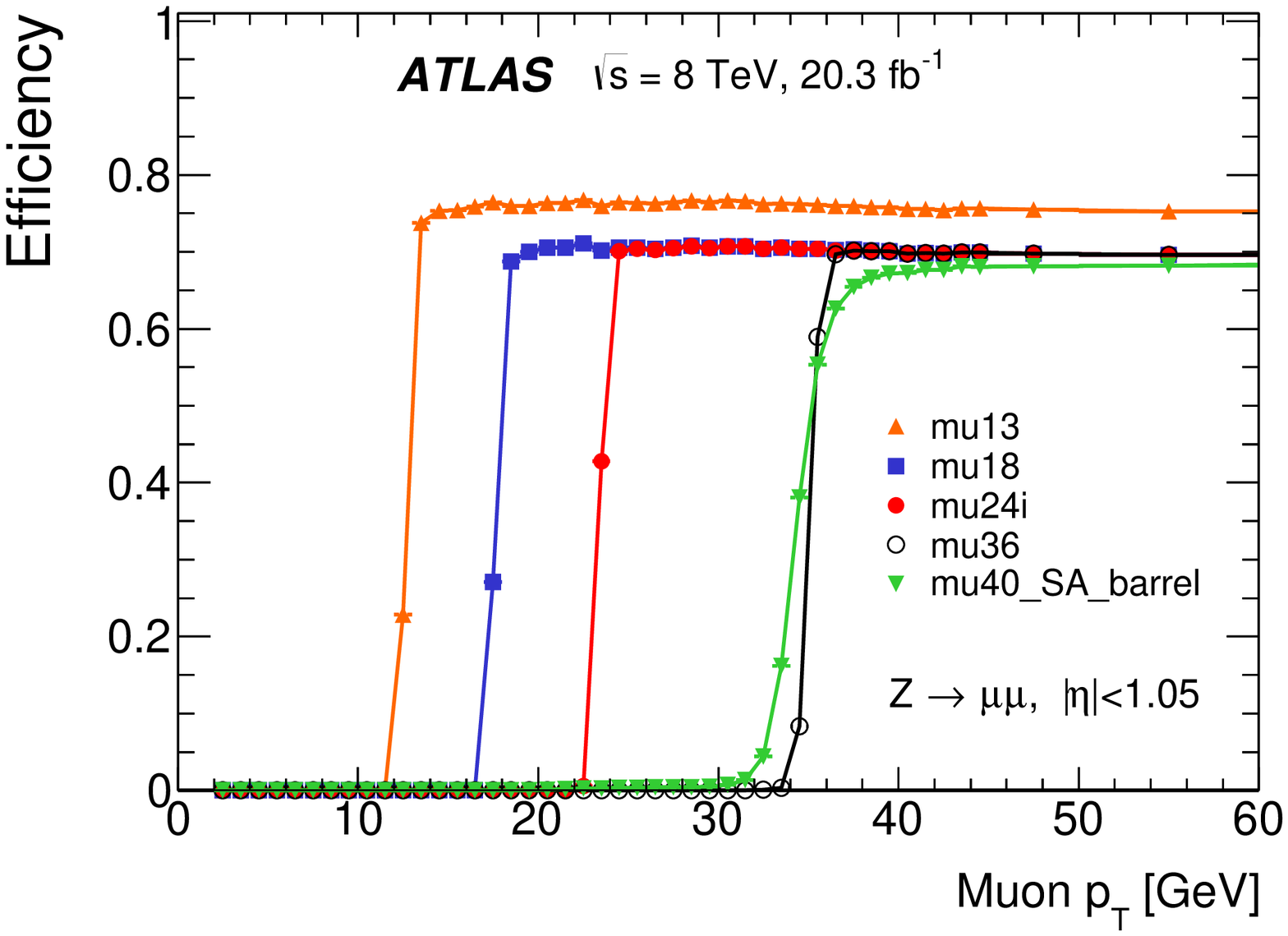}
    \label{fig:ztap_effi_mu18_data_barrel}
  }
\subfigure[]
  {\includegraphics[width=0.38\textwidth]{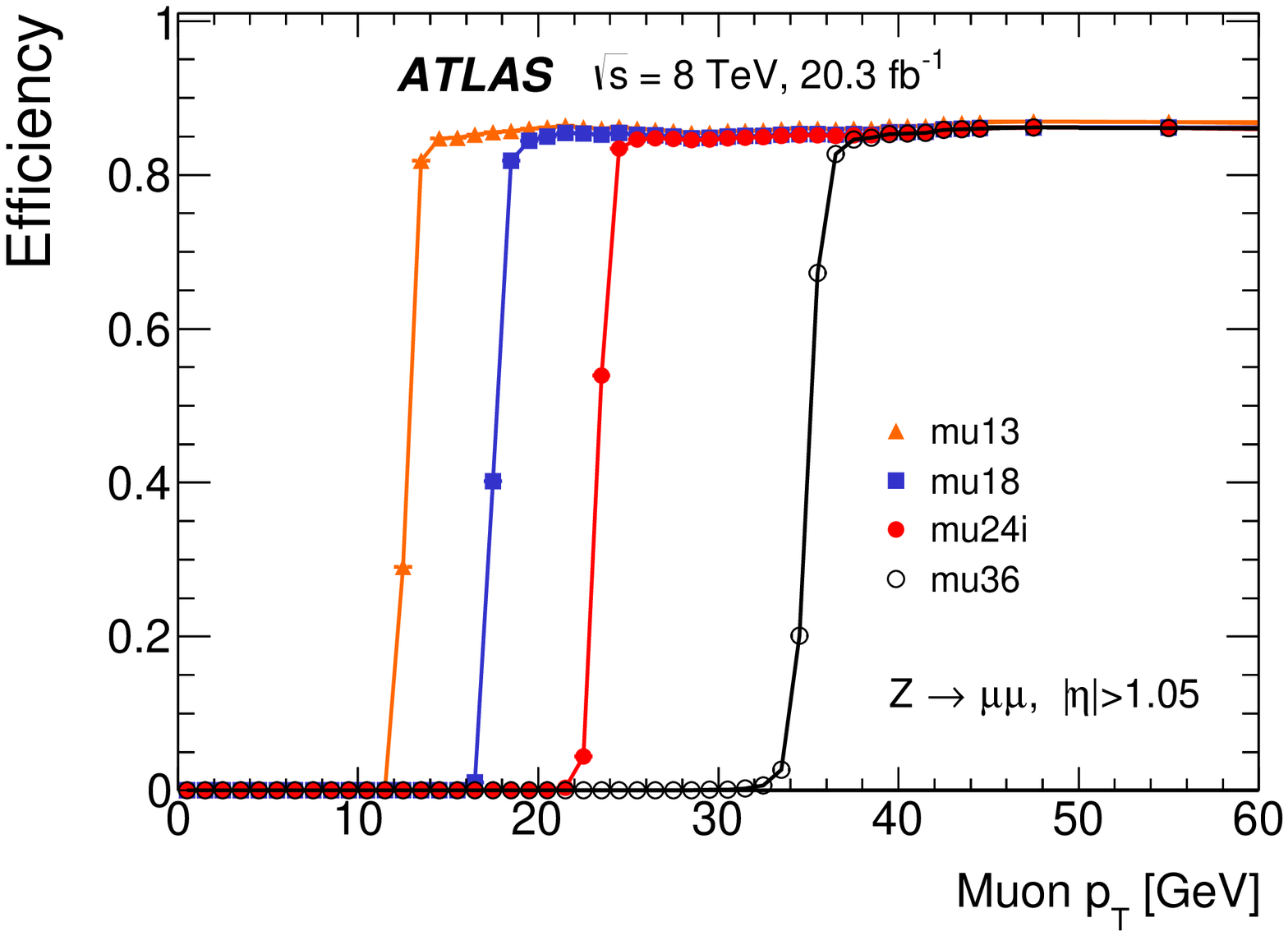}
  \label{fig:ztap_effi_mu18_data_endcap}
  }
\end{center}
\caption{Efficiency of single-muon triggers, 
mu13, mu18, mu24i, mu36 and mu40\_SA\_barrel,
measured in data as a function of the probe-muon transverse momentum {\pt},  for
(a) the barrel region and (b) the endcap region.
The error bars indicate statistical uncertainties only.}
\label{fig:ztap_effi_mu18}
\end{figure}

Figure~\ref{fig:ztap_effi_mu18} shows the efficiencies of
the mu36 trigger and of the mu40\_SA\_barrel trigger, together with that of mu24i trigger,
as measured in data.
%
The turn-on behaviour of mu24i and mu36 are sharp, while it is slower at
threshold for mu40\_SA\_barrel.
The latter relies only on the information from the muon detectors,
and thus the {\pt} resolution is coarser (see Sect.~\ref{sect:resolution}).
On the other hand, the requirement to pass either  mu36 or mu40\_SA\_barrel
results in about 2\,\% higher efficiency in the barrel region than achieved when requiring mu36 only,
because mu40\_SA\_barrel does not require an inner detector track match.
Therefore, requiring that  either the mu36 or mu40\_SA\_barrel triggers
are passed serves as a 
primary single-muon trigger for any processes that include muons with
$\pt \gtrsim 50$\,GeV.

Figure~\ref{fig:ztap_effi_mu18} also shows the efficiencies of
the medium-{\pt}, single-muon triggers, mu13 and mu18.
The plateau efficiency of mu13 is about 6\,\% higher in the barrel region
than that of mu18 and other higher- {\pt} triggers like mu24i.
This is because mu13 is seeded from Level-1 MU10, which requires a two-station coincidence, while
mu18 and the others are seeded from Level-1 MU15 which requires a three-station coincidence
(see Sect.~\ref{sect:trigger_l1}).

A fit using a Fermi function was performed  to 
quantify the turn-on behaviour of these medium-{\pt} single-muon triggers.
Table~\ref{tab:ztap_mu18_turnon_threshold} shows the evaluated plateau
and low edge {\pt} values for mu13 and mu18.
\begin{table*}
\caption{Result of Fermi function fit to the efficiency turn-on curve
  for the middle-{\pt} single-muon triggers.
The location  in {\pt} of the low edge of the plateau region is
  defined such that the efficiency decreases by 1\,\% from the plateau
  value. }
\label{tab:ztap_mu18_turnon_threshold}
\begin{center}
\begin{tabular*}{\textwidth}{@{\extracolsep{\fill}}llrrrr@{}}
\hline
               &               & \multicolumn{2}{c}{Data}   & \multicolumn{2}{c}{MC} \\
Trigger        &               & Plateau value & Low edge & Plateau value & Low edge \\
\hline
mu13     & Barrel        & 75.8\,\%        & 13.7\,GeV   & 75.0\,\%        & 12.8\,GeV  \\
               & Endcap        & 86.4\,\%        & 13.6\,GeV   & 86.1\,\%        & 13.4\,GeV  \\
mu18     & Barrel        & 70.1\,\%        & 18.2\,GeV   & 70.4\,\%        & 18.1\,GeV  \\
               & Endcap        & 85.7\,\%        & 18.7\,GeV   & 85.4\,\%        & 18.4\,GeV  \\ \hline
\end{tabular*}
\end{center}
\end{table*}
It is seen that the offline cut of muon $\pt > 15 (20)$\,GeV is sufficient
to ensure the mu13 (mu18) trigger efficiency is described by the plateau value.
These middle-{\pt} triggers are used in various triggers, such as
dimuon triggers 2mu13 and mu18\_mu8\_FS.
The efficiencies of the single-muon triggers, mu13 and mu18,
are necessary ingredients to calculate the dimuon trigger efficiencies.

\subsection{Full-scan-muon trigger}

\begin{figure}[tb]
\begin{center}
\subfigure[]
  {\includegraphics[width=0.38\textwidth]{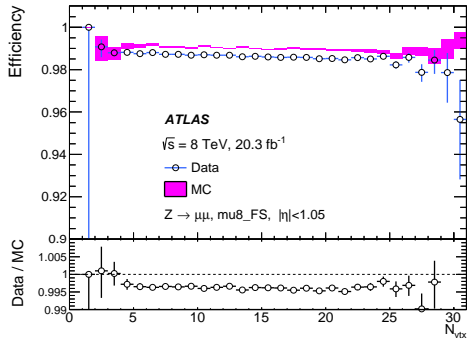}
  \label{fig:ztap_pileup_mu8fs_barrel}
  }
\subfigure[]
  {\includegraphics[width=0.38\textwidth]{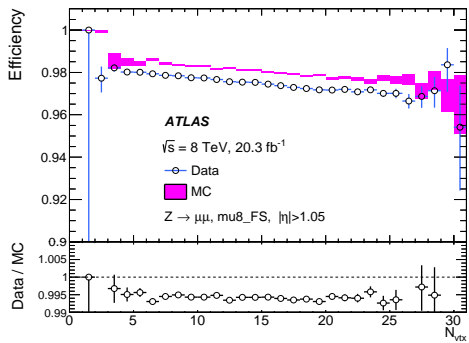}
  \label{fig:ztap_pileup_mu8fs_endcap}
  }
\end{center}
\caption{Efficiency of the mu8\_FS trigger measured as a function of the reconstructed number of
  vertices in an event, N$_{\mathrm{vtx}}$
   in (a) the barrel region and (b) the endcap region, in the data (dots) and in the MC simulation (bands)
  The lower panels show the ratio of efficiencies
  of data and MC simulation. The error bars represent statistical uncertainties only.}
\label{fig:ztap_pileup_mu8fs}
\end{figure}

As described in Sect.~\ref{sect:trigger_logic},
the  mu18\_mu8\_FS trigger is split into 
the RoI-based single-muon trigger, mu18,
and the full-scan triggers of mu18\_FS and mu8\_FS.
The full-scan trigger efficiencies were evaluated using the same method and 
sources of systematic uncertainties as for the single-muon trigger  (see Sect.~\ref{sect:ztap_syst}).
Only two sources of systematic uncertainties resulted in visible changes in the efficiency, while all others lead to negligible changes.
\begin{itemize}
\item Dependence on {\pt}:\\
  the uncertainty was estimated by comparing data and MC efficiencies as a function of {\pt}
  after correcting MC to reproduce data efficiency in $\eta$ and $\phi$.
  This resulted in a 0.2\,\% effect in the barrel and a 0.5\,\% effect in the endcap region.

\item Dependence on pile-up interactions:\\
  as shown in Fig.~\ref{fig:ztap_pileup_mu8fs},
  the efficiency has a small dependence on the number of pileup events
  in the end cap region,  with about 1.0\% efficiency loss per 20
  vertices. The MC simulation reproduces the effect well.
  This is accounted for by changing the distribution of the average number of
  pile-up interactions, resulting in a 0.1\,\% uncertainty.
\end{itemize}
The resulting uncertainties were added in quadrature to form the total systematic uncertainty.

\begin{figure}[tb]
\begin{center}
\subfigure[]
  {\includegraphics[width=0.38\textwidth]{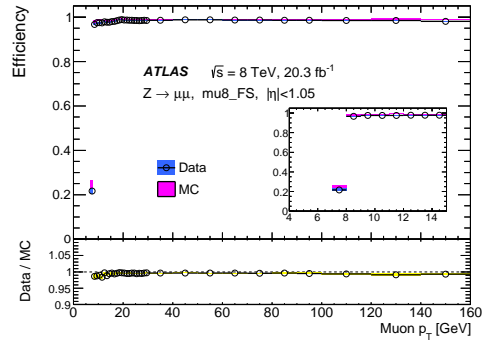}
  \label{fig:ztap_mu8fs_pt_barrel}
  }
\subfigure[]
  {\includegraphics[width=0.38\textwidth]{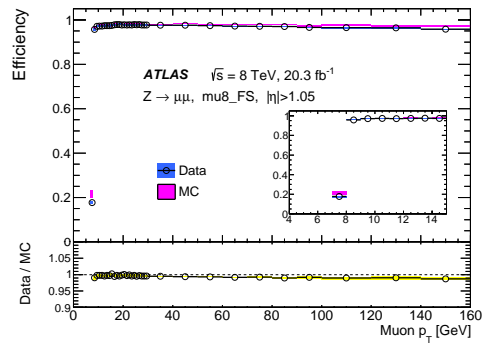}
  \label{fig:ztap_mu8fs_pt_endcap}
  }
\end{center}
\caption{Efficiency of the event-filter full-scan mu8\_FS as a function of the probe-muon transverse momentum {\pt},
separately in (a) the barrel region and (b) the endcap region.
}
\label{fig:ztap_eff_mu8fs_pt}
\end{figure}
Figure~\ref{fig:ztap_eff_mu8fs_pt} shows 
the data and MC efficiencies for the mu8\_FS trigger  for the barrel
and endcap regions. The efficiency plateaus for the barrel and endcap regions are
98.7\,\% and 97.6\,\%, respectively. This results in a higher efficiency for the dimuon trigger than achieved by requiring 
two RoI-based single-muon triggers.

\begin{figure}[tb]
\begin{center}
\subfigure[]
  {\includegraphics[width=0.35\textwidth]{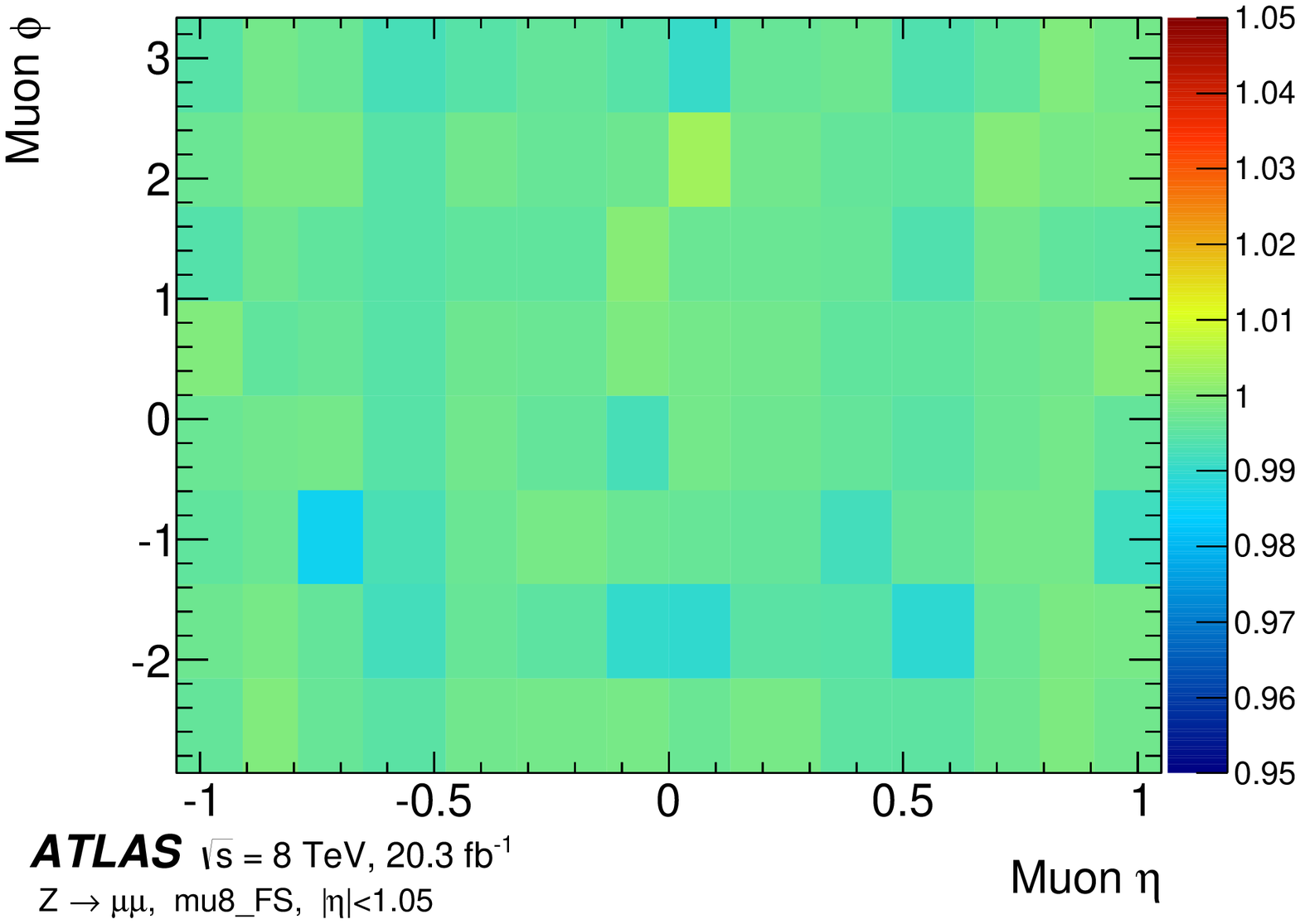}
  \label{fig:ztap_mu8fs_sf_barrel}
  }
\subfigure[]
  {\includegraphics[width=0.35\textwidth]{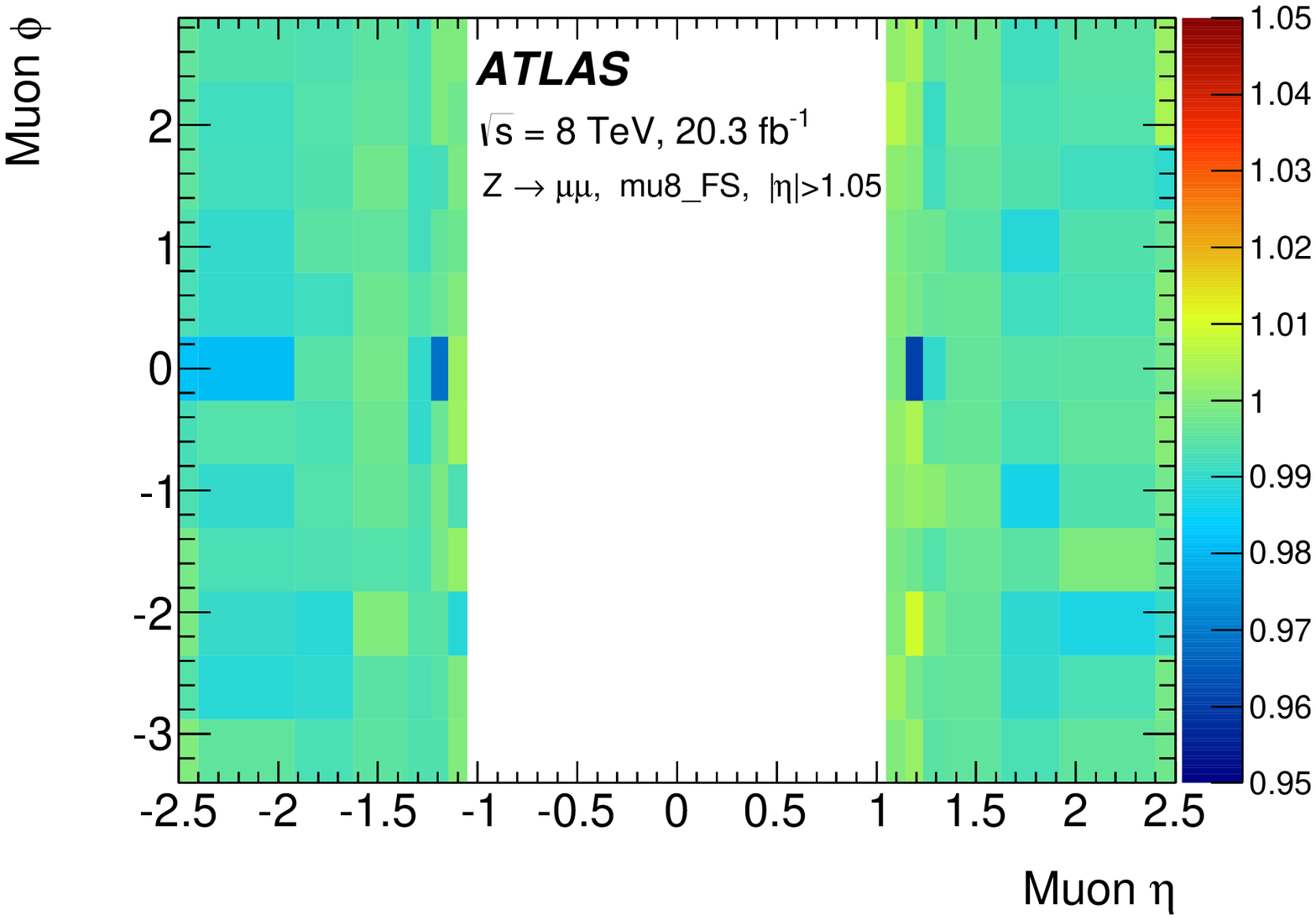}
  \label{fig:ztap_mu8fs_sf_endcap}
  }
\end{center}
\caption{Ratio of the data and MC efficiencies for the mu8\_FS trigger in bins of the probe-muon pseudorapidity $\eta$ and azimuthal angle $\phi$), 
 in (a) the barrel region and (b) the endcap region.}
\label{fig:ztap_eff_mu8fs_sf}
\end{figure}

 The ratio of the efficiencies in data and MC is shown as a function of $\eta$ and $\phi$ in Figure
~\ref{fig:ztap_eff_mu8fs_sf} for the probe-muons with $p_{T}$~10 GeV. It is consistent with unity to within 2\,\% except in two bins
where the difference is as large as 5\,\%.

%


%% file: jpsitap.tex
\section{Efficiency measurements at low \pt } 
\label{sect:jpsi_tap}

\subsection{Efficiency measurements with {\jpsi}}

For the kinematic region of $\pt \lesssim 10$\,GeV,
the efficiency was measured with the tag-and-probe method using {\jpsi} meson decays.

A MC study shows that the efficiency is slightly dependent on the measured $d_0$.
Therefore, the efficiencies of prompt and non-prompt muons can be different
due to different $d_0$ distribution.
This effect is mostly removed by 
the cuts on $d_0$ and $L_{xy}$ described in Sect.~\ref{sect:jpsi_tap_selection}.
The residual effect is then suppressed by reweighting the $d_0$ distribution
to that of the prompt muons, which is obtained from the events with $L_{xy}< 0$.

Owing to a very high purity of the offline muon identification,
the background also consists of muons, where the latter do not originate from the
decay of a  {\jpsi} meson.
The background fraction in the {\jpsi} mass window is about 16\,\%,
ranging between 13\,\% to 20\,\% depending on the muon {\pt}.
The efficiency was measured by correcting the background effect
using the side-bands of the invariant mass distribution.

\subsection{Systematic uncertainty}
\label{sect:jpsi_tap_syst}

The following sources of systematic uncertainty were evaluated.
The uncertainty numbers quoted in the following are for the
efficiency measured as a function of the probe-muon $\pt$ in the region of $4 < \pt < 10$\,GeV.
\begin{itemize}
\item Matching between probe-muon and trigger muon:\\
the effect was estimated by relaxing the 
$\Delta R$ criterion from 0.12 to 0.15, and also by relaxing the $\Delta R$ distance cut between the two muons
from 0.2 to 0.25.
The estimated uncertainty is up to 3\,\% (2\,\%) at  $\pt = 4$\,GeV in the barrel
(endcap) region, decreasing to 1\,\% at $\pt \gtrsim 6$\,GeV.

\item Reweighting of the $d_0$ distribution:\\
the effect was estimated by comparing the efficiency with that obtained by
not applying the $d_0$ reweighting.
The estimated uncertainty is 1\,\% at $\pt \sim 4$\,GeV,
decreasing to a negligible level at $\pt \gtrsim 6$\,GeV.

\item Probe-muon charge dependence:\\
the effect was estimated by comparing the efficiencies
measured with positively charged and with negatively charged probe-muons. The estimated uncertainty is 1\,\% at low $\pt \sim 4$\,GeV,
decreasing to 0.5\,\% at $\pt \gtrsim 6$\,GeV.

\item Background contribution:\\
the effect was estimated by not doing the background correction,
resulting in a uncertainty of 0.1\,\%.

\item Probe-muon selection criteria:\\
the effect was estimated by changing typically by 10\,\% the thresholds of various
selection criteria, leading to negligible effects.

\item Dependence on pile-up interactions:\\
Separate linear fits to the efficiency dependence on Nvtx in the data and MC simulation were performed in the range from Nvtx=5 to Nvtx=30 and extrapolated out to Nvtx=50.  The dependence on the fit range was observed to be negligible.  The largest difference between the fit results in data and MC simulation were observed to be 0.2 (0.4) \% in the barrel (endcap). This difference is taken as the estimate of the resulting  systematic uncertainty.
\end{itemize}

The total systematic uncertainties are obtained by adding 
the individual ones in quadrature.

\subsection{Low-{\pt} single-muon triggers}

\begin{figure}[tb]
\begin{center}
\subfigure[]
  {\includegraphics[width=0.38\textwidth]{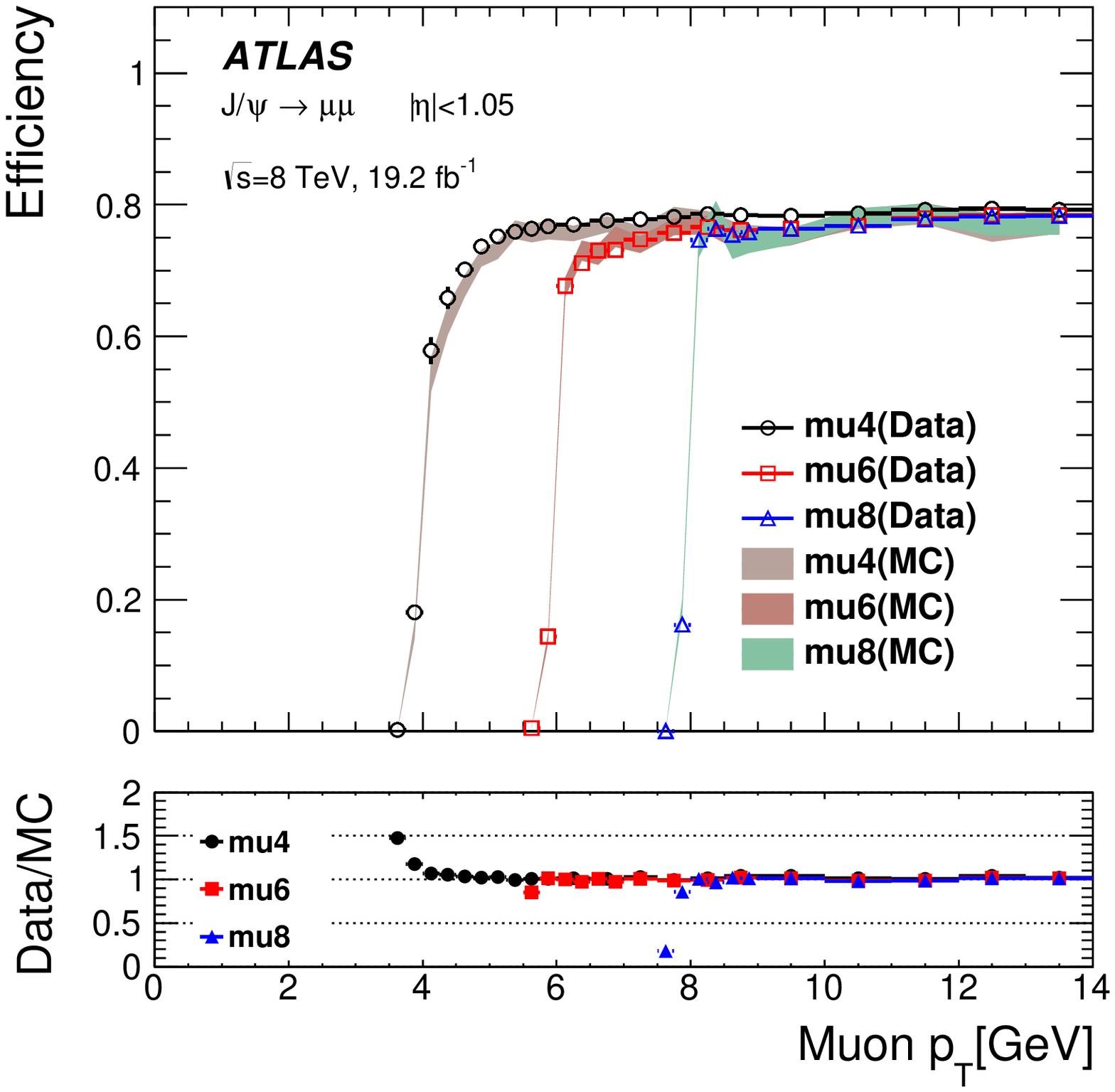}
    \label{fig:jpsieffi_mu468_barrel}
  }
\subfigure[]
  {\includegraphics[width=0.38\textwidth]{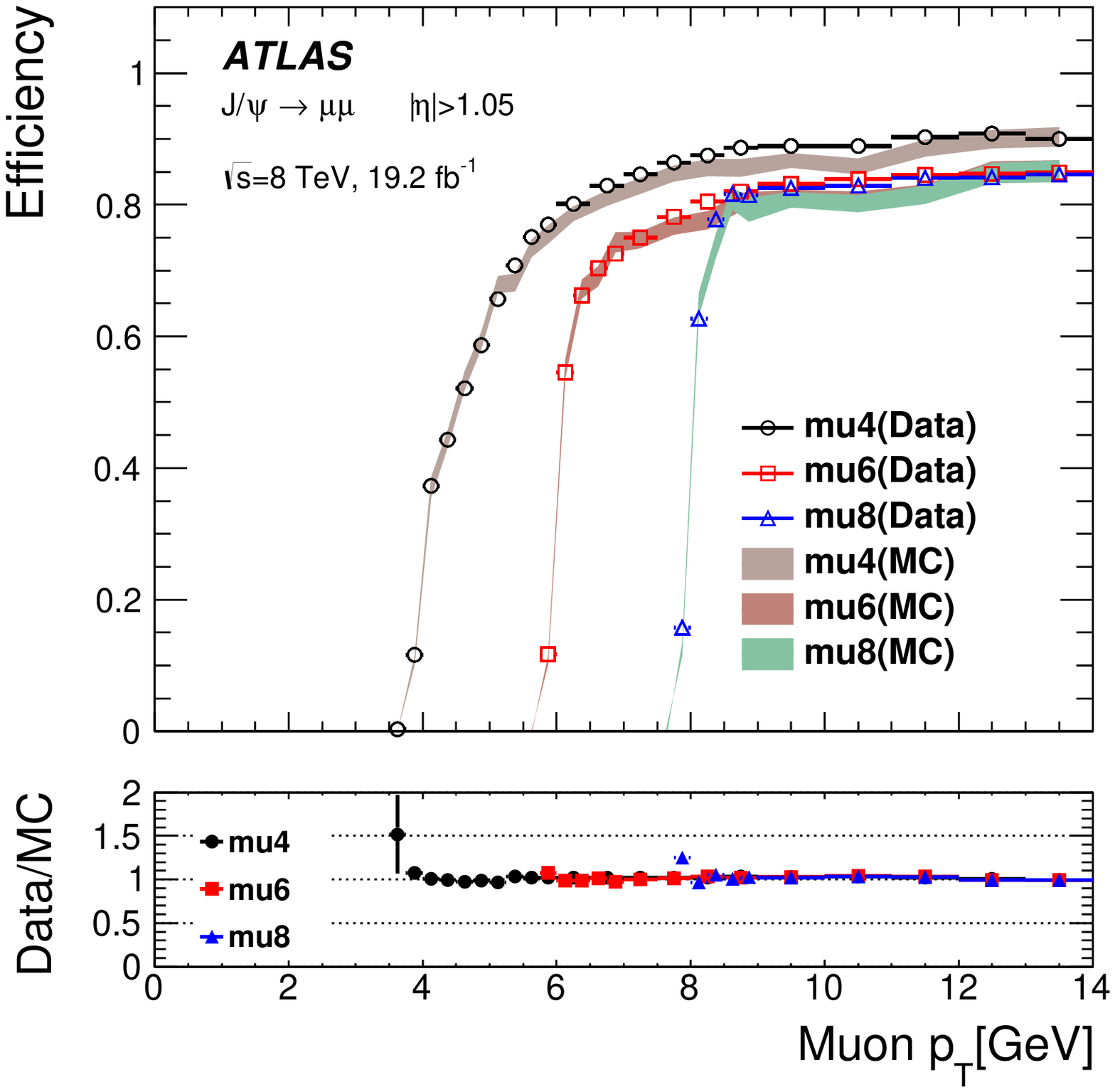}
  \label{fig:jpsieffi_mu468_endcap}
  }
\end{center}
\caption{Efficiency of low transverse momentun {\pt} single-muon triggers, mu4, mu6 and mu8,
   as a function of the probe-muon transverse momentum {\pt} in
  (a) the barrel region and (b) the endcap region, in the data (symbols) and in the MC (bands).
  The error bars for MC indicate the statistical uncertainties only, while those for data
  indicate both the statistical and systematic uncertainties.}
\label{fig:jpsieffi_mu468}
\end{figure}

Figure~\ref{fig:jpsieffi_mu468} shows the efficiency of the lowest-{\pt} single-muon
triggers, mu4, mu6 and mu8 as a function of the {\pt} of the probe-muon.
The efficiency of mu4 is about 40\,\% at the nominal threshold of 4\,GeV.
The mu4 turn-on curve rises slowly until $\pt \sim 8$\,GeV.
The plateau efficiency of mu4 is higher
by about 3\,\% in the endcap region, compared to
those of mu6 and mu8.

\begin{figure}[tb]
\begin{center}
\includegraphics[width=0.35\textwidth]{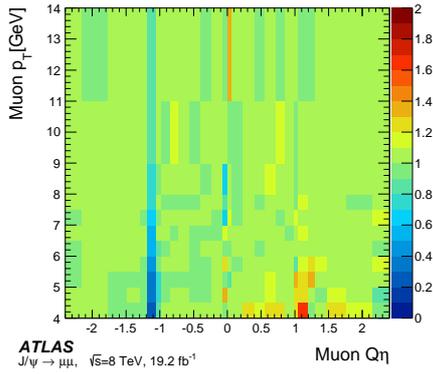}
\end{center}
\caption{Ratio of the data and MC efficiencies of the mu4 trigger in bins of the probe-muon
  pseudorapidity $\eta$ multiplied by its charge, $Q\eta$, and the transverse momentum {\pt}.}
\label{fig:jpsi_mu4_sf}
\end{figure}

The ratio of data and MC efficiencies
of the mu4 trigger  determined in bins of $\pt$ and $Q\eta$,
where $Q$ stands for the charge of the probe-muon is shown in
Fig.~\ref{fig:jpsi_mu4_sf}.
The ratio is significantly lower than unity at $Q\eta \sim -1.1$
for $\pt$  values up to $\sim 12$\,GeV.
In the muon spectrometer toroid magnetic field, the muons with $Q\eta >0$ ($<0$) bend toward the large (small) $\abseta$ direction in the $r$--$z$ plane.
The muons with $Q\eta \sim -1.1$ are thus likely to pass through only one layer of the RPC (see Fig.~\ref{fig:muon_schematic}) and hence are not triggered.
Figure~\ref{fig:jpsi_mu4_sf} shows that this is not well modelled in the
MC simulation. 


%% file: ttandw.tex
\section{Efficiency measurements at very high \pt} 
\label{sect:ttandw}

\subsection{Efficiency measurements with top quarks and {\Wboson} associated with jets}

For the kinematic region of $\pt \gtrsim 100$\,GeV,
the efficiency was measured using muons from top quark  and {\Wboson} +
jet candidate events.
Because they are statistically independent of each other
and also correspond to background-enriched samples of each other,
the efficiencies using muons in top quark and {\Wboson} + jet events
can be obtained by solving the following two equations 
\begin{eqnarray*}
\epsilon^{t, \rm data}
&=& f^{t,\rm data}_{t} \epsilon_{t} +  (1-f^{t,\rm data}_{t}) \epsilon_{\Wboson}, \\
\epsilon^{\Wboson, data}
&=& f^{\Wboson, \rm data}_{\Wboson} \epsilon_{\Wboson} + (1-f^{\Wboson, \rm data}_{\Wboson}) \epsilon_{t},
\end{eqnarray*}
where $\epsilon_{t(\Wboson)}$ is the efficiency in pure top quark ({\Wboson} + jets) events,
and $\epsilon^{t(\Wboson), \rm data}$ is the measured efficiency in
the top quark ({\Wboson} + jets) sample.
The factors $f^{t,data}_{t}$ and $f^{\Wboson,\rm data}_{\Wboson}$ denote the fraction
of true top quark ({\Wboson}+jets) events in the top quark  ({\Wboson} with jets) sample,
as determined by using MC simulation.

\subsection{Systematic uncertainty}

In the following, sources of systematic uncertainty are discussed and 
the  quoted uncertainty values are presented for the
efficiency measured using the {\Wboson} + jets sample
as a function of $\pt$, in the region of $100 < \pt < 400$\,GeV.
\begin{itemize}
\item  Muon isolation:\\
to estimate this effect, the efficiency was measured by varying 
the isolation cut, both by loosening and by tightening the criteria, as well as
by changing the $\Delta R$ cone size.
The estimated uncertainty is typically 0.2\,\%;

\item  Muon--jet separation:\\
the requirement on muon--jet separation serves also as an isolation cut.
This effect was estimated
by changing the $\Delta R$ criterion in the matching from 0.4 to 0.3 and 0.5.
The estimated uncertainty is typically 0.1\,\% and 0.3\,\% at maximum,

\item  {\MET} reconstruction:\\
the effect was estimated by changing the threshold from 20 GeV to 50 GeV, and also
by introducing another tight cut of $\MET \text{(calo)} > 120$\,GeV.
The estimated uncertainty is 0.5\,\% at maximum.

\item  Identification of $b$-jets:\\
the effect was estimated by repeating the measurements with a different
$b$-jet identification criterion, namely with 60\,\% efficiency and 80\,\% efficiency.
The estimated uncertainty is typically less than 0.1\,\%.

\item Cut on {\ptjet}:\\
the effect was estimated by raising the {\ptjet} threshold to 35\,GeV.
The estimated uncertainty is typically less than 0.1\,\%.

\item Background contribution:\\
the number of background events was estimated by using the dijet 
and {\Zboson} MC simulations and was found to be negligible at $\pt > 100$\,GeV.
\end{itemize}

All the contributions were added in quadrature to obtain
the total systematic uncertainties.

\subsection{Single-muon trigger efficiency at $\pt \gtrsim 100$\,GeV}

\begin{figure}[tb]
\begin{center}
\subfigure[]
  {\includegraphics[width=0.35\textwidth]{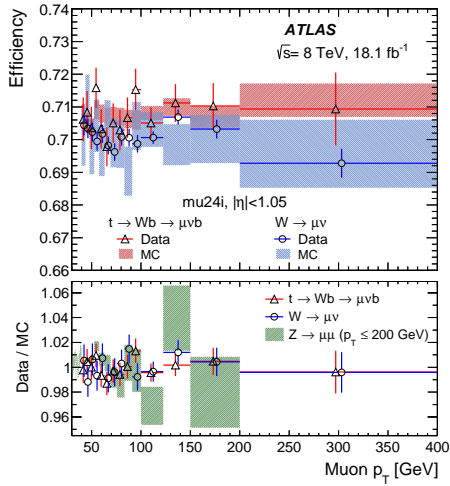}
  \label{fig:effi_mu24i_barrel}
  }
\subfigure[]
  {\includegraphics[width=0.35\textwidth]{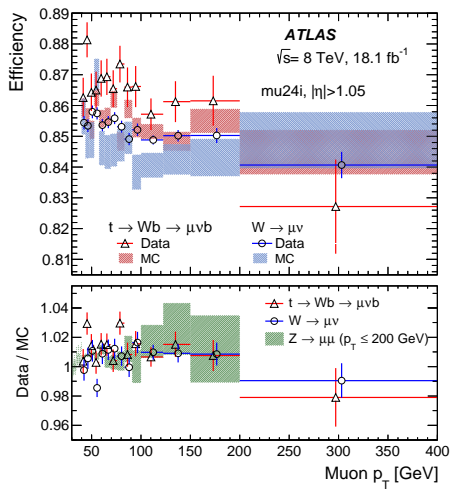}
  \label{fig:effi_mu24i_endcap}
  }
\end{center}
\caption{Efficiency of the mu24i trigger as a function of the probe-muon transverse momentum {\pt},
  as measured with the top quark and {\Wboson}+jet candidate events
  in the (a) barrel and (b) endcap regions.
  The lower panels show the ratio of the efficiencies
  in the data and MC simulation.
  Also shown is the efficiency as measured with the {\Zboson} decays using the
  tag-and-probe method.
  The error bars for MC simulation indicate the statistical
  uncertainties only while the error bars for data include both
  statistical and systematic uncertainties.
}
\label{fig:effi_mu24i}
\end{figure}

Figure~\ref{fig:effi_mu24i} shows the efficiencies measured
using top quark and {\Wboson} with jets events for the single isolated-muon trigger, mu24i,
in the barrel and endcap regions as a function of the {\pt} of the probe-muon,
up to $\pt \sim 400$\,GeV.
The data and MC simulation agree well up to the very high {\pt} values.

Also shown in Fig.~\ref{fig:effi_mu24i} are the ratios of the
efficiencies in the data and MC simulation for the three samples used for the efficiency determination. 
The three measurements are in good agreement with each other throughout a large {\pt} range,
providing a consistency check of the efficiency measurement in different physics processes
with different experimental techniques and in the presence of
different backgrounds. 

The efficiency in the end cap is seen to drop
off slightly at the highest $\pt$ which is not observed in the
barrel. This was further investigated and it was found that for the highest
energy muons ($\approx$ 1000 GeV) there is a slight loss of efficiency
at the event-filter when combining the muon spectrometer and inner detector track. While the offline algorithm looks for large energy deposits in the calorimeter which arise from bremsstrahlung, the event-filter algorithm always uses a paramterised energy loss for a minimum ionising particle. Without correction, this can cause a mismatch in the momentum estimate in the inner detector and muon spectrometer causing the combination to fail. This occurs in the end cap where kinematically, for fixed $\pt$, the energy of muons is much higher and thus high energy bremsstrahlung is more likely to occur. However, the effect is small, only occurs in the highest few {\pt} bins and accounts for a 4 $\%$ efficiency loss with a 2 $\%$ uncertainty.  


%% file: summary.tex
\section{Conclusions}

The ATLAS muon trigger has been successfully
adapted to the challenging environment at the LHC such that
stable and highly efficient data taking was achieved in
the year 2012.
The transverse momentum threshold for the single-muon trigger 
was kept at 24\,GeV, with a well-controlled trigger
rate of typically about 8.5\,kHz at the Level-1 and 65\,Hz at the
event-filter.
The processing times of the Level-2 and event-filter muon trigger algorithms
were sufficiently short to fit within the computing resource limitations.
The purity of the trigger is about 90\,\% at the event-filter, and more than half of the triggers originate
from electroweak bosons production.
The efficiencies are measured extensively
with the proton--proton collision data at a centre-of-mass energy
of 8\,TeV. 
The systematic uncertainty in the measured efficiency
for the single-muon trigger is evaluated to be about 0.6\,\% in a kinematic region of $25 < \pt < 100$\,GeV.
The efficiency was measured over a wide {\pt} range (few GeV to several hundred GeV)
by using muons from {\jpsi} mesons, {\Zboson}- and {\Wboson}-bosons, and
top quark decays showing highly uniform and stable performance. 


%% file: Acknowledgements.tex




We thank CERN for the very successful operation of the LHC, as well as the
support staff from our institutions without whom ATLAS could not be
operated efficiently.

We acknowledge the support of ANPCyT, Argentina; YerPhI, Armenia; ARC,
Australia; BMWFW and FWF, Austria; ANAS, Azerbaijan; SSTC, Belarus; CNPq and FAPESP,
Brazil; NSERC, NRC and CFI, Canada; CERN; CONICYT, Chile; CAS, MOST and NSFC,
China; COLCIENCIAS, Colombia; MSMT CR, MPO CR and VSC CR, Czech Republic;
DNRF, DNSRC and Lundbeck Foundation, Denmark; EPLANET, ERC and NSRF, European Union;
IN2P3-CNRS, CEA-DSM/IRFU, France; GNSF, Georgia; BMBF, DFG, HGF, MPG and AvH
Foundation, Germany; GSRT and NSRF, Greece; ISF, MINERVA, GIF, I-CORE and Benoziyo Center,
Israel; INFN, Italy; MEXT and JSPS, Japan; CNRST, Morocco; FOM and NWO,
Netherlands; BRF and RCN, Norway; MNiSW and NCN, Poland; GRICES and FCT, Portugal; MNE/IFA, Romania; MES of Russia and ROSATOM, Russian Federation; JINR; MSTD,
Serbia; MSSR, Slovakia; ARRS and MIZ\v{S}, Slovenia; DST/NRF, South Africa;
MINECO, Spain; SRC and Wallenberg Foundation, Sweden; SER, SNSF and Cantons of
Bern and Geneva, Switzerland; NSC, Taiwan; TAEK, Turkey; STFC, the Royal
Society and Leverhulme Trust, United Kingdom; DOE and NSF, United States of
America.

The crucial computing support from all WLCG partners is acknowledged
gratefully, in particular from CERN and the ATLAS Tier-1 facilities at
TRIUMF (Canada), NDGF (Denmark, Norway, Sweden), CC-IN2P3 (France),
KIT/GridKA (Germany), INFN-CNAF (Italy), NL-T1 (Netherlands), PIC (Spain),
ASGC (Taiwan), RAL (UK) and BNL (USA) and in the Tier-2 facilities
worldwide.

%% file: atlas_authlist.tex
\begin{flushleft}
{\Large The ATLAS Collaboration}

\bigskip

G.~Aad$^{\rm 84}$,
B.~Abbott$^{\rm 112}$,
J.~Abdallah$^{\rm 152}$,
S.~Abdel~Khalek$^{\rm 116}$,
O.~Abdinov$^{\rm 11}$,
R.~Aben$^{\rm 106}$,
B.~Abi$^{\rm 113}$,
M.~Abolins$^{\rm 89}$,
O.S.~AbouZeid$^{\rm 159}$,
H.~Abramowicz$^{\rm 154}$
% ...
\bigskip
\\
% ...
$^{11}$ Institute of Physics, Azerbaijan Academy of Sciences, Baku, Azerbaijan\\
% ...
$^{84}$ CPPM, Aix-Marseille Universit{\'e} and CNRS/IN2P3, Marseille, France\\
% ...
$^{89}$ Department of Physics and Astronomy, Michigan State University, East Lansing MI, United States of America\\
% ...
$^{106}$ Nikhef National Institute for Subatomic Physics and University of Amsterdam, Amsterdam, Netherlands\\
% ...
$^{112}$ Homer L. Dodge Department of Physics and Astronomy, University of Oklahoma, Norman OK, United States of America\\
$^{113}$ Department of Physics, Oklahoma State University, Stillwater OK, United States of America\\
% ...
$^{116}$ LAL, Universit{\'e} Paris-Sud and CNRS/IN2P3, Orsay, France\\
% ...
$^{152}$ Institute of Physics, Academia Sinica, Taipei, Taiwan\\
% ...
$^{154}$ Raymond and Beverly Sackler School of Physics and Astronomy, Tel Aviv University, Tel Aviv, Israel\\
% ...
$^{159}$ Department of Physics, University of Toronto, Toronto ON, Canada\\
% ...
\end{flushleft}
